\documentclass[pre,aps,onecolumn,showpacs,floatfix,10pt]{revtex4-1}
\usepackage[dvips]{graphicx}
\usepackage{amsmath}
\usepackage{amssymb}
\usepackage{color}

\begin{document}

\title{Temperature distribution in driven granular mixtures does not depend on mechanism of energy dissipation}
\author{Anna S. Bodrova, Alexander Osinsky and Nikolay Brilliantov}
\address{Skolkovo Institute of Science and Technology, 121205, Moscow, Russia}

\begin{abstract}
We study analytically and numerically the  distribution of granular temperatures in granular mixtures for  different dissipation mechanisms of inelastic inter-particle collisions. Both driven and force-free systems are analyzed.   We demonstrate that the simplified model of a constant restitution coefficient fails to predict even qualitatively a granular temperature distribution in a homogeneous cooling state. At the same time we reveal
 for driven systems   a stunning result -- the distribution of temperatures in granular mixtures is universal. That is, it does not depend on a particular dissipation mechanism of inter-particles collisions, provided the size  distributions of particles is steep enough. The results of the analytic theory are compared with simulation results obtained by the direct simulation Monte Carlo (DSMC). The agreement between the theory and simulations is perfect. The reported results may have important consequences for fundamental science as well as for numerous application, e.g. for the  experimental modelling in a lab of natural processes.
\end{abstract}
\maketitle
\section*{Introduction}
Mixtures of granular particles of different size are ubiquitous in nature and technology \cite{DryGranMed,GranRev,
PhysGranMed,SanPowGr}. The examples in nature range from pebbles and  sands to dust on the Earth. Many extraterrestrial
objects are  comprised of granular mixtures: One can mention interstellar dust clouds, protoplanetary discs
\cite{protodust} and planetary rings. Dense Saturn's rings demonstrate a tremendous size polydispersity, with the
particles size ranging from centimeters up to a few meters \cite{rings,pnas}.  Granular dust also covers the surface of
the Moon \cite{MoonSoil},  Mars \cite{mars} and possibly other planets and satellites. Industrial granular materials,
besides of pebbles and sands in building industry, are represented by powders in chemical and cosmetic production, as
well as salt, sugar and cereals in food industry.

Granular materials demonstrate very rich behavior -- depending on the applied load they can be in solid  liquid or
gaseous phase \cite{GranRev}. If the applied load is small, granular materials resist the external force and keep their
shape and volume as solids. With increasing external load they start to flow like fluids. Such unusual properties of
granular material stem from the dissipative nature of the inter-particle collisions, which are quantified by the so-called
restitution coefficient $\varepsilon$, see e.g.   \cite{GranRev,book}:
\begin{equation}
\label{rc} \varepsilon = \left|\frac{\left({\bf v}^{\,\prime}_{ki} \cdot {\bf e}\right)}{\left({\bf v}_{ki} \cdot {\bf
e}\right)}\right| \, .
\end{equation}
Here ${\bf v}^{\,\prime}_{ki}={\bf v}_{k}^{\,\prime}-{\bf v}_{i}^{\,\prime}$ and ${\bf v}_{ki}={\bf v}_{k}-{\bf
v}_{i}$ are the relative velocities of particles of masses $m_k$ and $m_i$ after and before a  collision, and ${\bf e}$ is a  unit vector directed along the inter-center vector  at the collision instant. 
The post-collision velocities  ${\bf v}_{k}^{\,\prime}$ and ${\bf v}_{i}^{\,\prime}$ are  related to the pre-collision velocities ${\bf v}_{k}$ and ${\bf v}_{i}$ as follows, e.g. \cite{book}:
\begin{equation}\label{v1v2} {\bf v}_{k/i}^{\,\prime} = {\bf v}_{k/i} \mp  \frac{m_{\rm eff}}{m_{k/i}}\left(1+\varepsilon\right)({\bf v}_{ki} \cdot {\bf e}){\bf e} \, .\end{equation}
Here $m_{\rm eff}=m_1m_2/\left(m_1+m_2\right)$ is the effective mass of colliding particles.
The restitution coefficient $0 \leq \varepsilon <1$ shows that the
after-collisional relative velocity is smaller than the pre-collisional one, since the mechanical energy is transformed 
into the internal degrees of freedom of the particles. Due to a permanent loss of the kinetic energy of particles in the 
collisions, a steady supply of energy is required to keep the system in liquid or gaseous phase (unless the
system is in a force-free state, where it undergoes a homogeneous cooling). The nature of the driving forces that
fluidize granular matter may be very different. These may be gravitational forces, as in the case of astrophysical
objects (dust clouds, protoplanetary disks and planetary rings) or avalanches in mountains \cite{av1}. It may be 
wind of atmospheric gases, initiating the motion of sand grains, which results in dune formation on the Earth
\cite{Dunes} or trigger dust storms on Mars \cite{mars}. The fluidization of granular materials in industry may be
caused by the vibration of a container, or by moving parts of a system, like e.g. blades or a piston.

The transport properties of granular fluids crucially depend on the mean kinetic energy of the grains, which is also
termed as "granular temperature". Due to dissipative nature of the inter-particles collisions, the energy
equipartition, valid for equilibrium molecular systems, does not hold for granular mixtures, where each species has its
own temperature. For a mixture of $i=1,2,\ldots N$ species the granular temperature of $k$-th species is defined as follows

\begin{equation}
\frac32 n_kT_k=\int d {\bf v}_k f_k\left({\bf v}_k, t\right)\frac{m_kv_k^2}{2}
\end{equation}
Here $m_k$ is the mass of the according granular species, ${\bf v}_k$ - its velocity, $f\left({\bf v}_k,t\right)$ - the
velocity distribution function, which quantifies the number of particles in the system of the kind $k$ with the
velocity $\textbf{v}_k$ at time $t$ and $n_k$ is the number density of the $k$-th species of the  granular fluid, $n_k= \int f_k\left({\bf v}_k,t\right) d {\bf
v}_k $. In what follows we consider granular fluids with a low density, which are termed as "granular gases". It is expected that the mixture behaves as a gas, when the total packing fraction of all components does not exceed about 20\%.

The violation of the energy equipartition in granular mixtures has been recognized almost two decades ago. It was
predicted theoretically, confirmed in computer simulations \cite{book,brey,dufty,GarzoDuftyMixture,Hrenya} and observed
experimentally \cite{wildman,menon}. An impressive natural example of a granular mixture with the broken energy
equipartition is Saturn rings, where all granular species demonstrate different temperatures
\cite{ringbook,Ohtsuki1999,Ohtsuki2006,Salo1992b}. The polydispersity in the rings arises due to coagulation and fragmentation of granular particles\cite{pnas, frank2004}. Although the effect of broken equipartition is known for a long
time, still the physical laws that determine the distribution of granular temperatures in granular mixtures are not
known. Such laws should predict the granular temperature for each species as a function of (i) size and mass
distribution of granular particles, (ii) of the dissipative mechanism of the particles collisions and (iii) of the
driving mechanism, applied to the system to keep it fluidized. Force-free granular mixtures can exist in a gaseous
state in the regime of homogeneous cooling; here the temperature distribution should be determined by the items (i) and
(ii) above. The temperature distribution in granular mixtures with the simplified model of a constant restitution
coefficient, $\varepsilon= \rm const$, which is equal for all inter-particle collisions, has been addressed in our
previous study \cite{lev}; some universality of the granular temperature distribution was reported \cite{lev}. Although
the assumption of a constant restitution coefficient drastically simplifies the analysis and is widely used, see e.g.
\cite{h83,gz93,dp03,dppre03,n03,nebo79,neb98,ap06,ap07}, it contradicts, the experimental observations
\cite{w60,bhd84,kk87}, as well as basic mechanical laws \cite{rpbs99,titt91}. The latter indicate  that $\varepsilon$
does depend on the impact velocity \cite{rpbs99,kk87,bshp96,mo97,sp98}. This dependence may be obtained by solving the
equations of motion for colliding particles with the explicit account for the dissipative forces acting between the
grains. The simplest, but still rigorous, first-principle model of inelastic collisions takes into account the
viscoelastic properties of particles' material. This results in the corresponding inter-particle forces
\cite{bshp96,goldobin} and eventually, in the restitution coefficient for viscoelastic particles
\cite{rpbs99,sp98,delayed}:
\begin{equation}
\label{epsx} \varepsilon_{ki} = 1 + \sum^{20}_{j=1} h_j \left( A \kappa_{ki}^{2/5} \right)^{j/2} \left |\left
({\bf{v}}_{ki}\cdot\textbf{e}\right )\right |^{j/10}, \qquad \qquad \kappa_{ki}= \kappa\,
\frac{i+k}{i^{5/6}k^{5/6}\sqrt{i^{1/3}+k^{1/3}}},
\end{equation}
where $h_k$ are numerical coefficients, $\kappa$ and $A$ characterize respectively the elastic and dissipative properties of the particles material (see Methods).  Viscoelastic model agrees well with the experimental
data when the impact velocity is not very large \cite{kk87,Falcon1,Labous,Falcon2}. If the dissipative mechanism is caused
by plastic deformation of particles, one obtains the following expression for the restitution coefficient
\cite{Thornton}:
\begin{equation}\label{eplastic}
\varepsilon=a \left( 1-b^2/6 \right) \left(1+ 4 \sqrt{3/5} \sqrt{ b^{-2} -6^{-2}} \right)^{1/4} \qquad a= \left(6
\sqrt{3}/5 \right)^{1/2} \qquad b= \frac{V_{\rm yield}}{\left(\textbf{e}\cdot\textbf{v}_{ki}\right)},
\end{equation}
where $V_{\rm yield}$ is the yield velocity. Dissipative mechanism associated with plastic deformation corresponds to rather high impact velocities \cite{Labous,Falcon2}. A phenomenological exponential model for the velocity-dependent
restitution coefficient has been also employed for the description of experimental data in Ref. \cite{LunSavage},
\begin{equation}\label{eLun}
\varepsilon=\exp\left(-\delta\sqrt{\frac{\rho}{Y}}\left(\textbf{e}\cdot\textbf{v}_{ki}\right)\right)\,\,,
\end{equation}
where $\delta$ is a dimensionless parameter, $\rho$ is the density of the particle material and $Y$ is the Young modulus.

It is well known that the velocity dependence of the restitution coefficient may drastically change the qualitative
behavior of granular systems. For instance, it changes the cooling law in a homogeneous cooling state
\cite{rpbs99,briltemp}, the velocity distribution function \cite{briltemp,annapre}, the diffusion of granular particles
\cite{NBTPSelfDif2000} and even pattern formation, which becomes a transient process for the case of velocity-dependent
$\varepsilon$ \cite{ClustersPRL2004}. Therefore, to formulate the laws for the temperature distribution in a granular
mixture one needs to consider in detail the dissipation mechanism of particles collision. This is done in the present
study. We analyze the distribution of granular temperatures in mixtures of granular gases for different dissipative
mechanisms and driving models. We show that the simplified model of a constant restitution coefficient fails to predict
even qualitatively the granular temperature distribution in a homogeneous cooling state. At the same time for driven
granular systems we arrive at an astonishing result -- the distribution of temperatures in granular mixtures
is universal, that is, it does not depend on a particular dissipation mechanism of particles collisions. This
conclusion holds true for steep distributions of particles size. The results of the analytic theory of the
present study are compared with simulation results obtained by the direct simulation Monte Carlo (DSMC). The agreement
between the theory and simulations is perfect.

\section*{Results and discussion}
\subsection{Model}

We consider a polydisperse granular system with discrete distribution of masses of particles. Let the smallest particle
mass be $m_1=0.01$ and masses of other particles read, $m_k=km_1$, where $k=1,2,...N$ are integers and $N$ is the
total number of different species in the system. The system is spatially uniform and dilute enough so that only pairwise collisions take place in the system and multiple collisions of the particles may be safely neglected. The mass-velocity
distribution function $f_k(\textbf{v}_k,t)$ gives concentrations of particles of mass $m_k$ with the velocity $\textbf{v}_k$
at time $t$. Since the deviation of the velocity distribution function from the Maxwellian distribution is relatively small \cite{book}, we assume for
simplicity that $f_k(\textbf{v}_k,t)$ is Maxwellian. This function evolves according to the Boltzmann equation, applicable for granular gases, where correlations of velocities of colliding particles may be neglected \cite{book},
\begin{equation}
\label{eq:BEgen} \frac{\partial}{\partial t} f_k\left(\textbf{v}_k, t \right) =  \sum_{i=1}^{N} I^{\rm coll}_{ki} + I^{\rm heat}_{k}.
\end{equation}
In Eq. \eqref{eq:BEgen} $I^{\rm coll}_{k}$ is the Boltzmann collision integral \cite{book}:
\begin{equation}
I^{\rm coll}_{ki} =\sigma_{ki}^2\int d{\bf v}_{i} \int d{\bf e} \, \Theta (-{\bf v}_{ki}\cdot {\bf e}\,)\left|{\bf v}_{ki} \cdot {\bf e}\, \right|\,  \,
\left[\chi f_k({\bf v}_{k}^{\ \prime\prime},t)f_i ({\bf v}_{i}^{\ \prime\prime},t)-f_k({\bf v}_{k},t)f_i({\bf
v}_{i},t)\right] \label{II},
\end{equation}
where $\sigma_{ki}=\left(\sigma_k+\sigma_i\right)/2$, with $\sigma_{k}=\left(6m_k/(\pi\rho)\right)^{1/3}$ being the
diameter of particles of mass $m_k$, $\rho$ is the mass density of the particle material. 
The summation is performed over all species in the system. ${\bf v}_{k}^{\, \prime\prime}$ and ${\bf v}_{i}^{\, \prime\prime}$ are  pre-collision
velocities in the so-called inverse collision, resulting in the post-collision velocities ${\bf v}_{k}$ and ${\bf
v}_{i}$.   The Heaviside function $\Theta(-{\bf v}_{ki}\cdot{\bf e})$ selects the approaching particles and the factor
$\chi$ equals the product of the Jacobian of the transformation $\left({\bf v}_k^{\,\prime\prime}, \, {\bf
v}_i^{\,\prime\prime}\right) \to \left({\bf v}_{k}, \, {\bf v}_{i}\right)$ and the ratio of the lengths of the
collision cylinders of the inverse and the direct collisions \cite{book}:
\begin{equation}
\label{Jacobdef} \chi = \frac{\left|{\bf v}_{ki}^{\ \prime\prime}\cdot{\bf e}\right|}{\left|{\bf v}_{ki}\cdot{\bf
e}\right|} \, \frac{{\cal D}\left({\bf v}_k^{\ \prime\prime},{\bf v}_i^{\ \prime\prime} \right)} {{\cal D}\left({\bf
v}_k,{\bf v}_i\right)}
\end{equation}
In the case of a constant restitution coefficient $\chi=1/\varepsilon^2$ for viscoelastic particles it has a more
complicated form \cite{book}.

The second term $I^{\rm heat}_k$ describes the driving of the system. It quantifies the energy injection into a
granular gas to compensate its losses in dissipative collisions; it is zero for a gas in  a homogeneous cooling state
(HCS). Here we consider a uniform heating -- the case when the grains suffer small random uncorrelated  kicks
throughout the volume \cite{WilliamsMacKintosh1996,Sant2000}. To mimic the external driving forces a few types of
thermostat have been proposed \cite{Sant2000}. For a thermostat with a Gaussian white noise, the heating term has the
form \cite{vne98,Sant2000}:
\begin{equation}
I^{\rm heat}_k=\frac12\frac{\Gamma_k}{m_k}\frac{\partial^2}{\partial\bf v_k^2}{f_k}({\bf v_k},t).
\end{equation}
Here the constant $\Gamma_k$ characterizes the strength of the driving force. It may vary for different species,
depending on the type of driving \cite{zippelius}. When all species are supplied with the same energy, we have a
driving equipartition,  $\Gamma_k=\Gamma_1=\rm const.$  In the case of the force controlled driving $\Gamma_k \propto
1/m_k$, while in the case of the velocity controlled driving $\Gamma_k \propto m_k$ \cite{zippelius}. In our study we
analyze a more general case of a power-law dependence of $\Gamma_k$ on a particle mass, namely, $\Gamma_k = \Gamma_1
k^{\gamma}$. The driving may also depend on the local velocity of granular particles \cite{Cafiero}, however we neglect
this effect in the present study.

Multiplying the Boltzmann equation (\ref{eq:BEgen}) by $m_kv_k^2/2$ for $k=1...N$ and performing the integration over
${\bf v}_k$, we get the following system of equations for evolution of the granular temperatures of species of
different masses:
\begin{eqnarray}
\frac{dT_1}{dt} = -T_1\sum_{i=1}^{N}\xi_{1i} +\Gamma_k \cr \ldots\cr \frac{dT_k}{dt} = -T_k\sum_{i=1}^{N}\xi_{ki}
+\Gamma_k\cr \ldots\cr \frac{dT_N}{dt} = -T_N\sum_{i=1}^{N}\xi_{Ni} +\Gamma_k. \label{sys}
\end{eqnarray}
In a homogeneous cooling state $\Gamma_k=0$ and the granular system permanently cools down. Driven granular systems,
that is, systems with a thermostat rapidly settle into a non-equilibrium steady state and all granular temperatures
attain after a short time, some constant values, so that $dT_k/dt=0$ and the above system (\ref{sys}) turns into a set
of algebraic equations,
\begin{equation}\label{Tsteady}
T_k\sum_{i=1}^{N}\xi_{ki} =\Gamma_1k^{\gamma}.
\end{equation}
For the constant restitution coefficient the cooling rate $\xi_{ki}$, quantifying the decrease of granular temperature of species of mass $m_k$ due to collisions with species of mass $m_i$ is given by the following expression \cite{lev}:
\begin{equation}
\xi_{ki}(t) =\frac{8}{3}\sqrt{2\pi}n_i\sigma_{ki}^{2}g_{2}(\sigma_{ki})
\left(\frac{T_km_i+T_im_k}{m_im_k}\right)^{1/2}\left(1+ \varepsilon_{ki}\right) \left(\frac{m_i}{m_i+m_k}\right)\left[1-\frac{1}{2}\left(1+ \varepsilon_{ki}
\right)\frac{T_im_k+T_km_i}{T_k\left(m_i+m_k\right)} \right] \label{xikconst}.
\end{equation}
In Ref. \cite{lev} it was assumed that $\varepsilon_{ki} =\varepsilon$. In the case of viscoelastic particles the
cooling rates may be also computed and the result reads (see Methods for detail):
\begin{equation}
\xi_{ki}(t) =
\frac{16}{3}\sqrt{2\pi}n_i\sigma_{ki}^{2}g_{2}(\sigma_{ik})
\left(\frac{T_km_i+T_im_k}{m_im_k}\right)^{1/2}\!\left(\frac{m_i}{m_i+m_k}\right)
\left[1-\frac{T_km_i+T_im_k}{T_k\left(m_i+m_k\right)}+\sum_{n=2}^{\infty}B_n\left(h_n-\frac12\frac{T_km_i+T_im_k}{T_k\left(m_i+m_k\right)}A_n\right) \right]
\label{xikvisc}
\end{equation}
where $A_n=4h_n+\sum_{j+k=n}h_jh_k$ are pure numbers and
\begin{equation}
B_n(t)\!=\left(A\kappa_{ki}^{2/5}\right)^{\frac{n}{2}}\!\left(2\frac{T_km_i+T_im_k}{m_im_k}\right)^{\frac{n}{20}}\!\left(\frac{\left(20+n\right)n}{800}\right)\!\Gamma\left(\frac{n}{20}\right)
\end{equation}
with $\Gamma\left(x\right)$ being the Gamma-function.

While it is possible to obtain the explicit expressions for the cooling coefficients $\xi_{ki}$ for the viscoelastic
dissipative model, it is not the case for the  elasto-plastic model. Therefore it is practical to exploit a notion
of a "quasi-constant" restitution coefficient, which corresponds to the effective restitution coefficient averaged over
all collisions; it depends on the current temperature of granular fluid \cite{annapre}, but not on the impact velocity.

\subsection{Effective restitution coefficient of colliding particles in a granular mixture}

The equations (\ref{epsx}), (\ref{eplastic}) and  (\ref{eLun}) for the restitution coefficient $\varepsilon$ give this
quantity for a collision with a particular impact velocity. Since a wide range of the impact velocities is observed,
one deals with a wide range of restitution coefficients. Naturally, this is much more complicated than to deal with a
single number, of a simplified model of a constant restitution coefficient. Therefore for practical reasons it is worth
to define a restitution coefficient $ \left< \varepsilon_{ki} \right>$, averaged over all possible collisions. Such
effective "quasi-constant" restitution coefficient may be defined as follows \cite{annapre},
\begin{equation}
\label{colav} \left\langle \varepsilon_{ki}\right\rangle = \frac{ \int d {\bf v}_k d {\bf v}_i d {\bf e} f({\bf
v}_k,t)f({\bf v}_i,t)\Theta\left(-v_n\right) \left| v_n \right| \varepsilon_{ki} \left(v_n\right) } {\left\langle
\left|v_n\right|\right\rangle}.
\end{equation}
Here $v_n=\left({\bf v}_{ki} \cdot {\bf e}\right)$ is the normal component of the relative velocity of particles and $\left \langle \left|v_n\right|\right\rangle$ is its collisional average:
\begin{equation}
\left \langle \left|v_n\right|\right\rangle=2\sqrt{2\pi}\sqrt{\frac{T_k}{m_k}+\frac{T_i}{m_i}}
\end{equation}

The Heaviside function $\Theta\left(-v_n\right)$ selects the approaching particles. This quantity has been introduced and
tested in \cite{annapre} for a uniform one-component granular gas and demonstrated its adequacy. Here we generalize
this concept for a mixture of granular particles of different masses $m_i$ and granular temperatures $T_i$. Using again
the Maxwellian approximation for the velocity distribution function, the collisional average yields for the effective
coefficient of the viscoelastic particles (see Methods for detail):

\begin{equation}
\left\langle \varepsilon_{ki}\right\rangle = 1+{\sum^{20}_{n=1}}\frac{2^{1+\frac{n}{20}}}{2+\frac{n}{10}} \Gamma
\left(2+\frac{n}{20} \right)h_n\left(A\kappa_{ki}^{2/5}\right)^{n/2}\left(\frac{T_k}{m_k}+\frac{T_i}{m_i}\right)^{n/20}.
\label{epseff}
\end{equation}
In contrast to the case of one-component granular gas \cite{annapre}, the effective restitution coefficient depends on
the granular temperatures of both components $T_k$ and $T_i$, but not on the impact velocity, since it characterizes
the collisions in average.  The "quasi-constant" restitution coefficient (\ref{epseff}) may be now used in Eq.
(\ref{xikconst}) for the cooling rates in place of the constant restitution coefficients; that is, the following
substitute may be used $\varepsilon_{ki} \to \left\langle \varepsilon_{ki}\right\rangle$. Although Eq. (\ref{epseff})
refers to viscoelastic particles, one apply the collisional averaging (\ref{colav}) for other dissipation models. 

\subsection{Homogeneous cooling state: The failure of simplified model of constant $\varepsilon$ }

\begin{figure}\centerline{\includegraphics[width=0.6\textwidth]{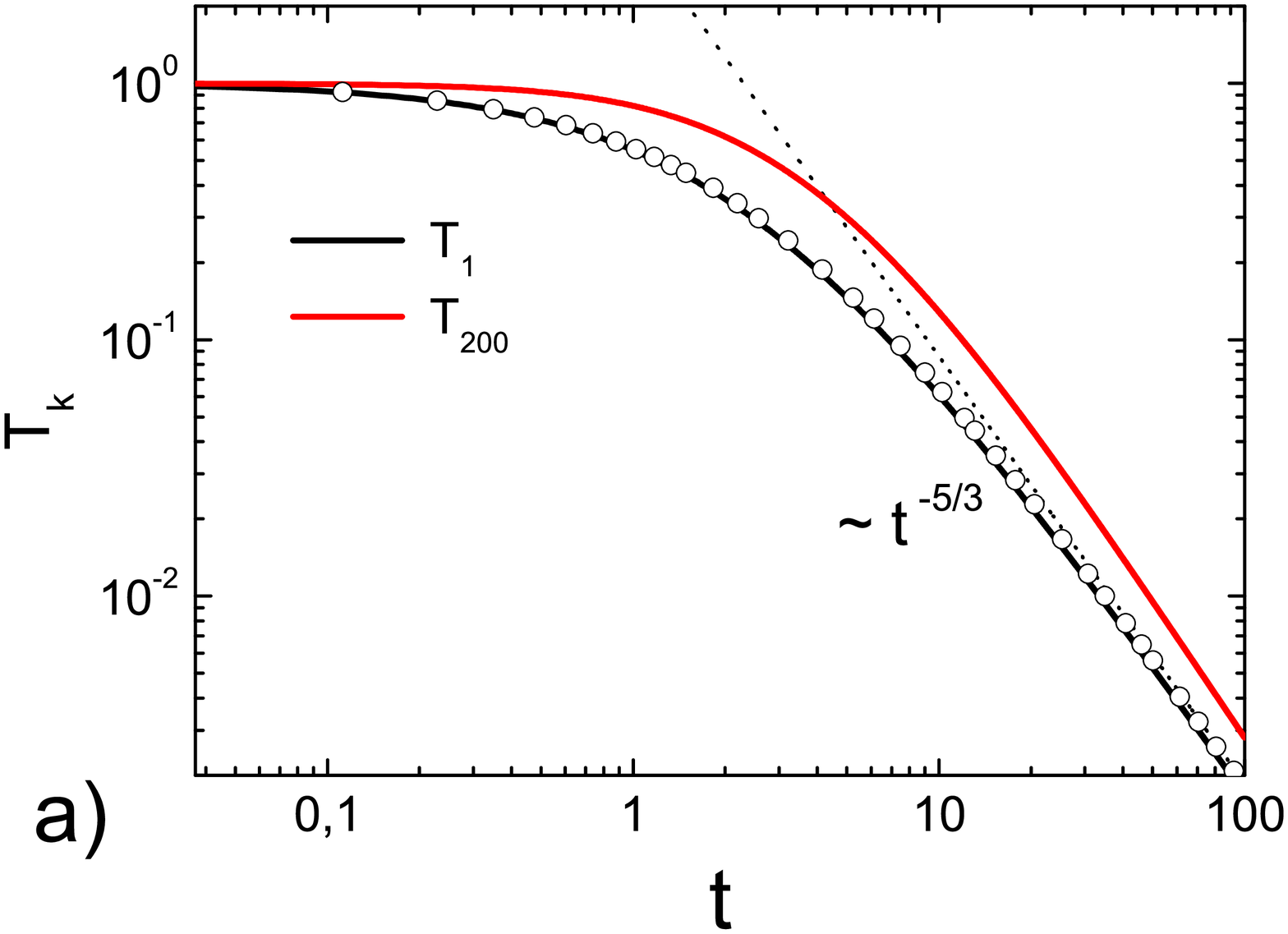}\includegraphics[width=0.6\textwidth]{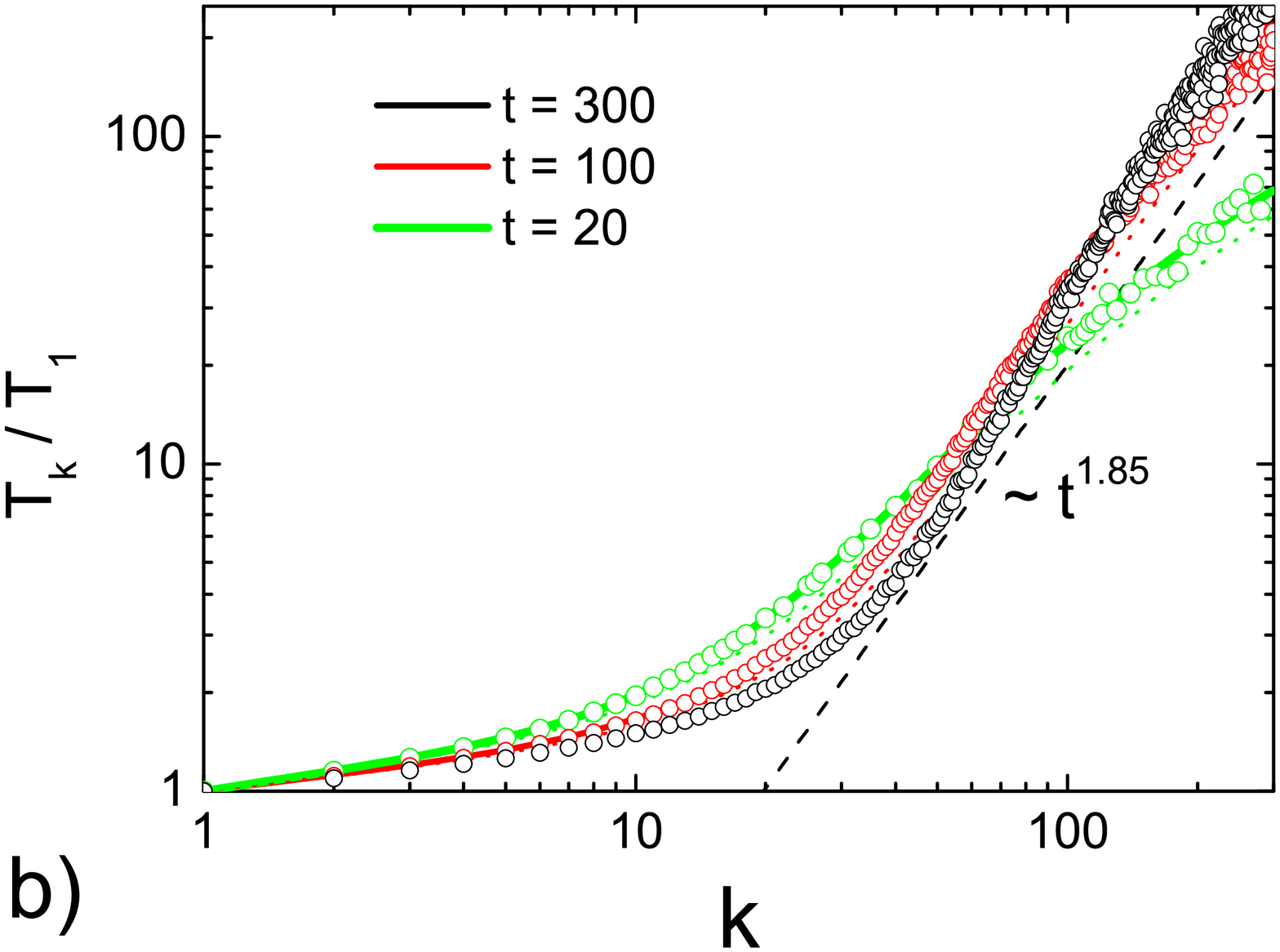}}
\caption{a) Evolution of granular temperatures $T_1$ and $T_{200}$ in the granular gas in a homogeneous cooling statefor the velocity-dependent restitution coefficient, Eq. (\ref{epsx}). The dotted line shows the asymptotics $\sim t^{-5/3}$. Symbols show the results of the DSMC simulations. b) Dependence of the granular temperatures $T_k$ in the HCS on the reduced mass of the particle $k=m_k/m_1$ at
different times. Solid lines correspond to viscoelastic particles (solution of system of equations (\ref{sys}) with
$\xi_{ik}$ in the form  Eq.~(\ref{xikvisc}) with  $A\kappa^{2/5}=0.441$.   Symbols show the results of the DSMC simulations. The dotted line shows the slope of the temperature distribution for the case of a constant $\varepsilon$.} \label{GAHCS}
\end{figure}

\begin{figure}\centerline{\includegraphics[width=0.6\textwidth]{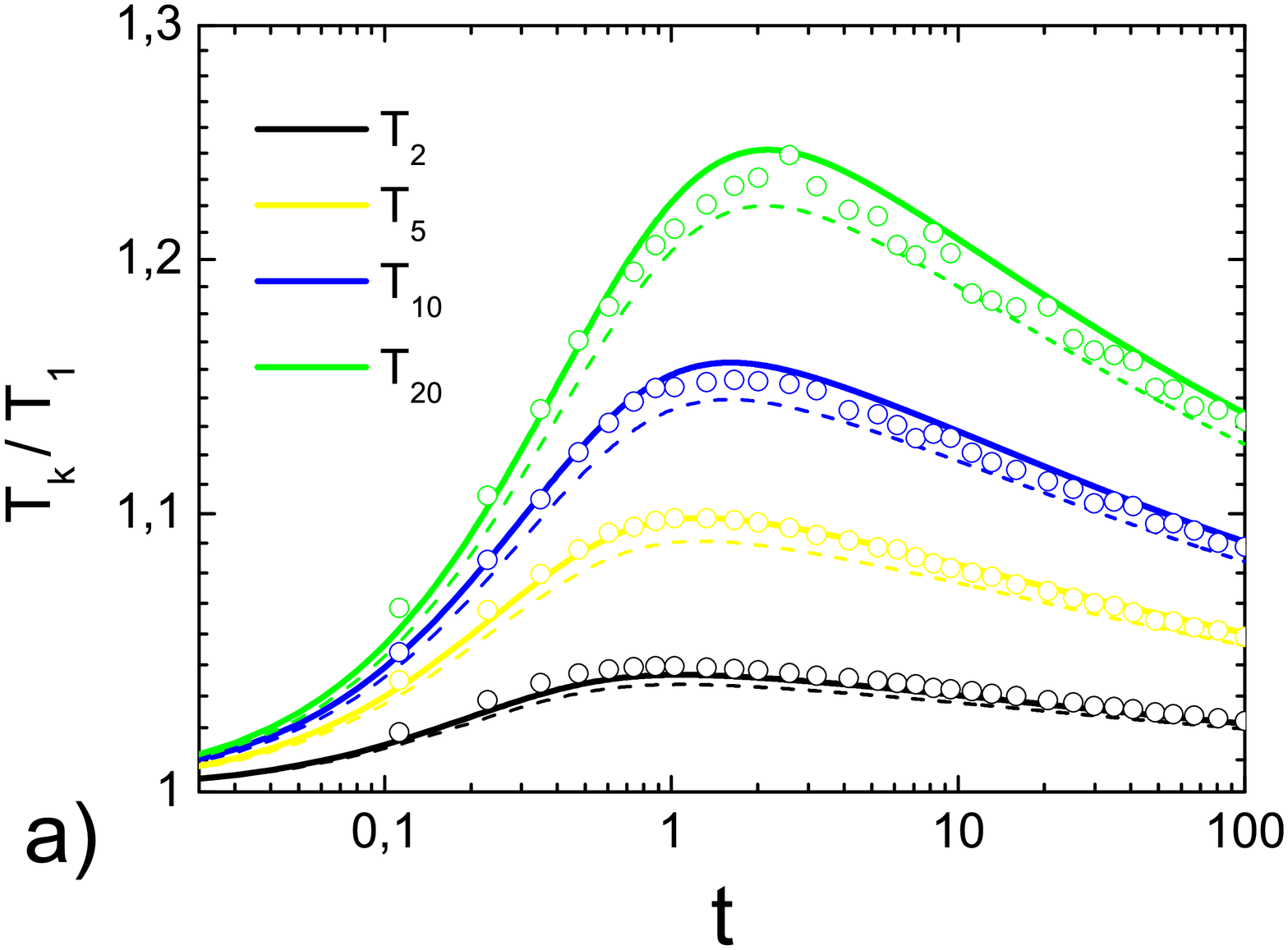}\includegraphics[width=0.6\textwidth]{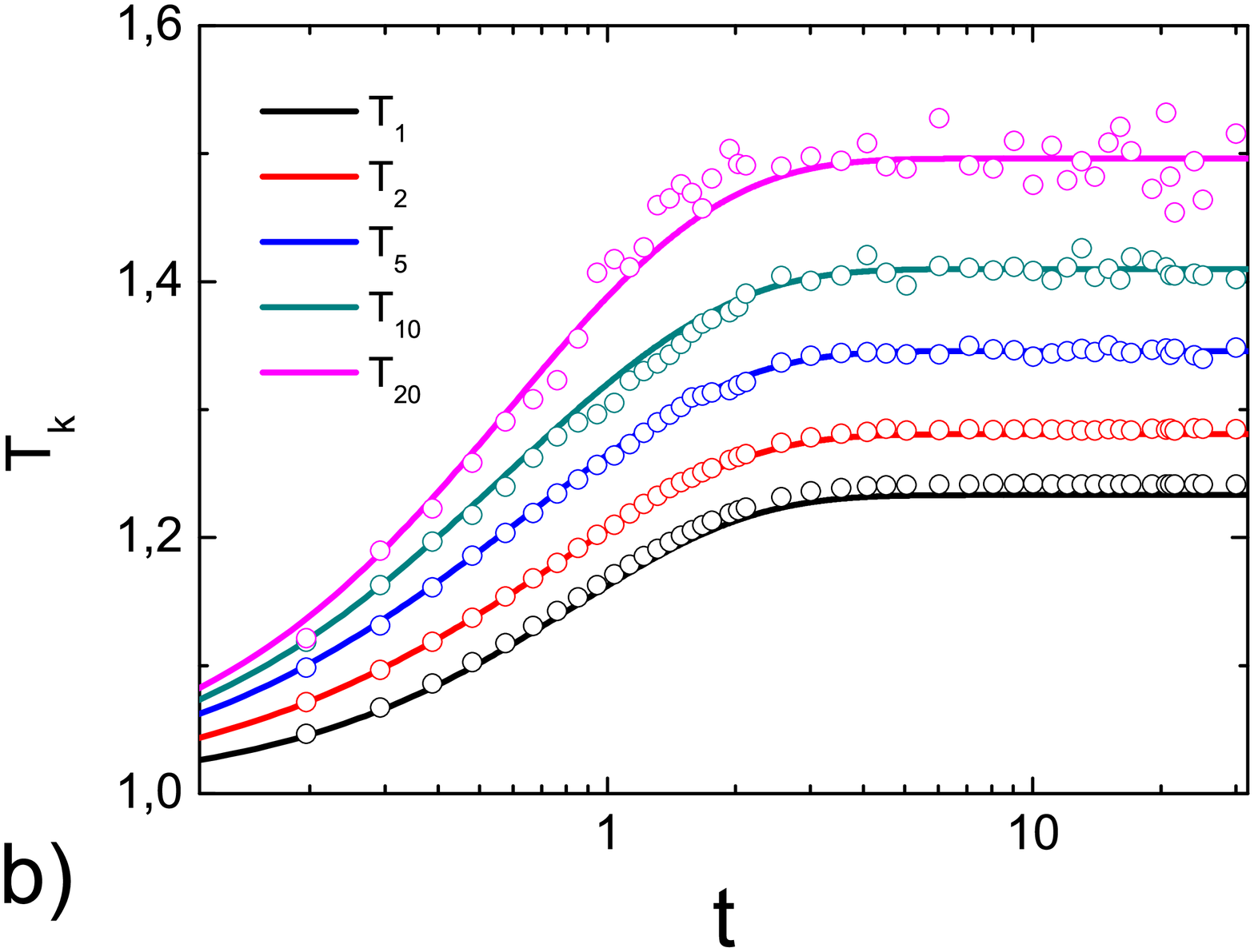}}\caption{Evolution of granular temperatures $T_k$. a) In a homogeneous cooling state. Solid lines illustrate the direct solution of the  system of equations (\ref{sys}) with $\xi_{ki}$ in the form of (\ref{xikvisc}). Dashed lines show the evolution of $T_k/T_1$ of particles, colliding with an effective restitution coefficient (\ref{epseff}) with $A\kappa^{2/5}=0.063$. b) In a uniformly heated granular gas with $\Gamma_k=\Gamma_1=1$ (all species are supplied with the same energy) for $A\kappa^{2/5}=0.063$. The notations are the same as in the panel (a). } \label{GTG1visco}\end{figure}

In order to calculate the granular temperatures of particles of different masses in the homogeneous cooling state we
have  solved numerically the system of differential equations (\ref{sys}) with zero heating rate, $\Gamma_k=0$ and
cooling coefficients $\xi_{ki}$ (Eqs. \ref{xikvisc}), corresponding to viscoelastic particles. We exploit the size
distributions of particles, which is steep enough, $n_k \simeq k^{-\theta}$ with $\theta=3$. The evolution of granular
temperatures $T_k(t)$ is shown in Fig.~\ref{GAHCS}a. The behavior of a granular gas of viscoelastic particles is
drastically different as compared to the granular gas of particles colliding with a constant restitution coefficient.
While in the case of constant restitution coefficient the granular temperature decrease with the same rate,
corresponding to the Haff's law \cite{book}, the evolution of granular temperatures of viscoelastic particles is rather
complicated. The temperature of monomers $T_1$ of mass $m_1$ cools down according to $t^{-5/3}$ (the generalized Haff's
law \cite{book}), while the temperature of more massive particles decrease slower at the initial state of cooling
(retarded cooling) and faster at the later state of cooling (accelerated cooling) (Fig.~\ref{GAHCS}a). The difference
becomes more pronounced with increasing mass, as it is shown in Fig.~\ref{GTG1visco}a, where the evolution of the ratios of granular temperatures is shown.
In the case of a constant restitution coefficient the ratios of all granular temperatures tend to the steady state
\cite{annapre}, while for viscoelastic particles the ratio of granular temperatures, $T_k/T_1$, first grows, then reaches
its  maximal value and then decreases with time,  tending  to unity, that is, tending to the equipartition  (Fig.~\ref{GTG1visco}a).  The larger the mass $m_i$ of the particle, the larger the granular temperature and at the
later time the maximum of $T_k/T_1$ is achieved. The temperature distribution $T_k/T_1$ evolves with time and changes
its form, see Fig.~\ref{GAHCS}b. It does not correspond to the distribution $T_k \sim  k^{1.85}$, observed for the
system with a constant restitution coefficient \cite{lev}. Hence the behavior of granular temperatures with a realistic
restitution coefficient qualitatively differs from that predicted for  a model with  a constant $\varepsilon$. In
other words the simplified model of a constant restitution coefficient, widely used in the scientific literature, fails to
describe a complicated behavior of a granular mixture. At the same time, as it is follows from  Fig.~\ref{GTG1visco}a, the
application of the corresponding quasi-constant, temperature-dependent restitution coefficient (\ref{epseff}) allows to
model a granular mixture with an acceptable accuracy. Figs. \ref{GAHCS} and \ref{GTG1visco} also demonstrate that the
theoretical results obtained by the solution of the rate equations (\ref{sys}) are in a perfect agreement with  the numerical simulations by the DSMC (see Methods for more detail).

\begin{figure}\centerline{\includegraphics[width=0.6\textwidth]{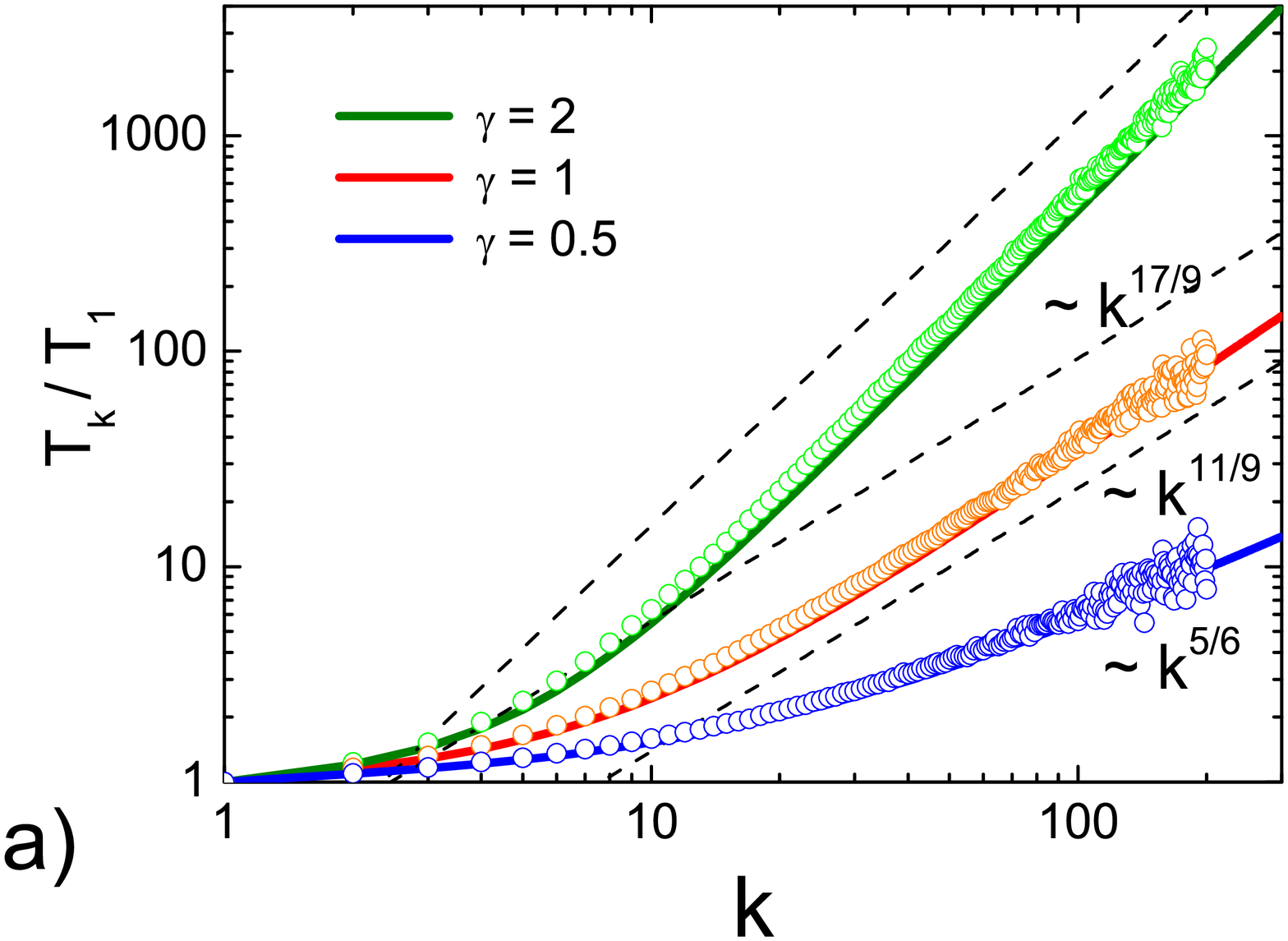}\includegraphics[width=0.6\textwidth]{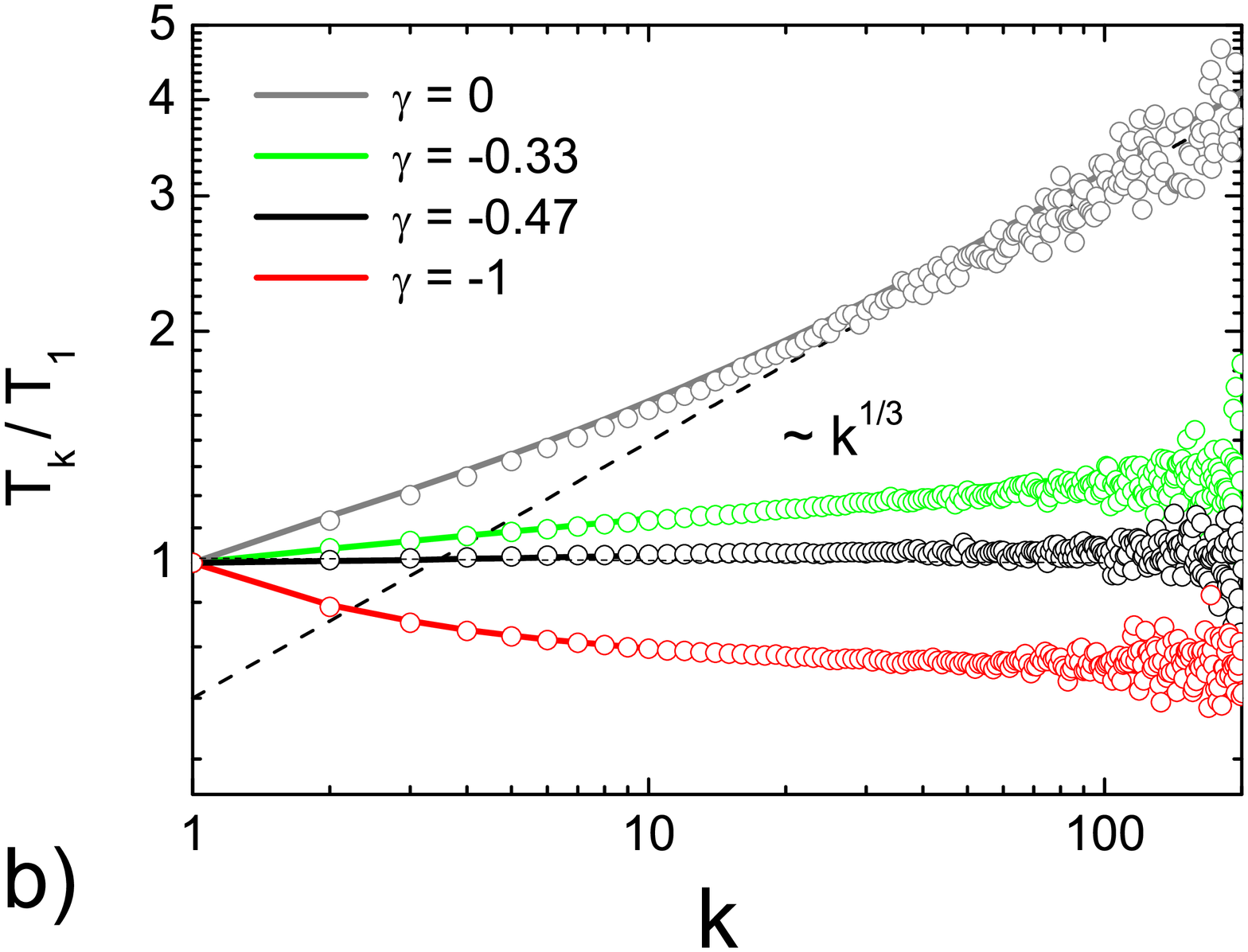}}
\caption{ Dependence of the granular temperatures $T_k$ on the reduced mass of the particle $k=m_k/m_1$ for a heated
granular gas with $\Gamma_k=k^{\gamma}$. a) For $\gamma=1$ (velocity controlled driving), $\gamma=2$ and $\gamma=0.5$. The results of the solution of the system with the effective restitution coefficient are shown with thin lines (cooling rates $\xi_{ki}$, Eq.~(\ref{xikconst}) with the restitution coefficient $\left\langle
\varepsilon_{ki}\right\rangle$, Eq.~(\ref{epseff})), while the full solution of the system is given by the thick lines (cooling rates $\xi_{ki}$, Eq.~(\ref{xikvisc})). Two solutions are practically indistinguishable. Symbols show the results of the DSMC simulations. The viscoelastic parameter is $A\kappa^{2/5}=0.063$. b) For $\gamma=0$ (equal distribution for the energy input for all species) and negative values of the exponent $\gamma$:
$\gamma=-0.33, -0.47, -1$. The notations are the same as for the panel (a). The viscoelastic parameter is $A\kappa^{2/5}=0.441$.  } \label{GAG1}
\end{figure}

\subsection{Driven granular mixtures: Universality of temperature distribution for all dissipative mechanisms}
To study the evolution of a driven granular mixture we solve the system of equations (\ref{sys}) with the
mass-dependent heating rates $\Gamma_k=\Gamma_1k^{\gamma}$. We start from the case of viscoelastic particles and use
the cooling coefficients (\ref{xikvisc}) and then the cooling coefficients (\ref{xikconst}) for the impact-velocity
independent restitution coefficients $\varepsilon_{ki}$, with the use of the effective quasi-constant
coefficients, $\left\langle \varepsilon_{ki} \right \rangle$ from Eq. (\ref{epseff}) in the place of $\varepsilon_{ki}$. The evolution  of granular
temperatures in a heated gas is shown at Fig.~\ref{GTG1visco}b. As one can see from the figure, all temperatures relax
to the steady-state values. The steady state temperatures $T_k$ form a stationary  distribution, that
behaves for large $k$ as a power law, see Fig.~\ref{GAG1}.

To understand this behavior theoretically, we assume the power-law distribution for the steady state temperatures,
\begin{equation}
T_k=T_1k^{\alpha},
\end{equation}
provided $k$ are not small. Approximating for $N\gg 1$ the summation by integration in Eq.~(\ref{Tsteady}) with the cooling
rates, Eqs. (\ref{xikvisc}), we get:
\begin{align}\label{int0}
\sum_{i=1}^{N}\xi_{ki} & =c  \int_{1}^{N}  di n_i \left(i^{1/3}+k^{1/3}\right)^{2}\frac{i}{k+i} \left(i^{\alpha-1} \!
+k^{\alpha-1} \right)^{1/2}\times\left[1-\frac{i}{k+i} \left(1+\left(\frac{i}{k}\right)^{\alpha-1}\right)+\right.
\\\nonumber
 & + \left.\sum_{n=2}^{\infty}\left(A\kappa_{ik}^{2/5}\right)^{\frac{n}{2}}\!\left(2\left(i^{\alpha-1} \! +k^{\alpha-1} \right)\right)^{\frac{n}{20}}\!\left(\frac{\left(20+n\right)n}{800}\right)\!\Gamma\left(\frac{n}{20}\right)\left(h_n-\frac12 A_n\frac{i}{k+i}
\left(1+\left(\frac{i}{k}\right)^{\alpha-1}\right)\right) \right]
\end{align}
with
\begin{equation}
\label{xi0} c=\frac{4}{3}\sqrt{2\pi}\sigma_1^2 \left(\frac{T_1}{m_1}\right)^{\frac{1}{2}}\,.
\end{equation}
If the particle size-distribution $n_i=n_i(i)$ is steep enough the main contribution to the integral in
Eq.~(\ref{int0}) comes from $i \ll k$. Expanding the integrand in  Eq.~(\ref{int0}) with respect to $(i/k) \ll 1$ and
keeping only the leading terms in the expansion we arrive at (see Methods for more detail):
\begin{equation}
\label{eq:xi_gen} \sum_{i=1}^{N}\xi_{ki}   = \left\{
    \begin{array}{ll}
         ck^{\frac{\alpha}{2}- \frac56} \int_1^N i\, n_i \, di & \mbox{if } \, \, \alpha \geq 1 \\
         {} \\
         ck^{- \frac13} \int_1^N i^{\frac{\alpha +1}{2}} \, n_i \, di & \mbox{if } \, \, 0< \alpha < 1 \, .
    \end{array}
\right.
\end{equation}
Here we exclude $\alpha <0$, since it may yield for $i \ll k$ a negative sign for the factor in the square brackets of
the integrand in Eq.~(\ref{int0}). The result of the integration, however, should be positive, as it gives the cooling
rate. For steep distributions $n_i$ one can approximate $N$ in the upper limits of the integrals in
Eq.~(\ref{eq:xi_gen}) by the infinity,
\begin{equation}
\label{eq:steep_cond} \int_1^N i^{p} \, n_i \, di \simeq \int_1^{\infty} i^p\, n_i \, di =\rm const \,,
\end{equation}
where $p=1$ for $\alpha>1$ and $ p=(\alpha +1)/2 $  for $0< \alpha <1$ (see Eq.~(\ref{eq:xi_gen})), so that the sum in
(\ref{Tsteady}) does not (asymptotically, for $N \gg 1$) depend on $N$. Taking into account that the exponents of $k$
in the both sides of Eq. (\ref{Tsteady}) must be equal, we finally arrive at:
\begin{equation}
\label{eq:alpha_heat} \alpha    = \left\{
    \begin{array}{ll}
         \frac{5}{9} + \frac{2}{3}\gamma & \mbox{if } \qquad  \gamma \geq \frac23 \\
         {} \\
         \gamma + \frac{1}{3} & \mbox{if } \qquad  -\frac13 \leq \gamma \leq \frac23.
    \end{array}
\right.
\end{equation}
Surprisingly, this result exactly coincides with the one for the velocity-independent restitution coefficient
$\varepsilon_{ik} = \varepsilon ={\rm const.}$ of Ref. \cite{lev}.

In Fig.~\ref{GAG1} the theoretical predictions for the temperature distribution are compared with the results of the
numerical solution of equations (\ref{sys}) with cooling rates of viscoelastic particles, with the cooling rates for
the effective quasi-constant restitution coefficient, as well as  with the DSMC results (see Methods for the application
detail of the DSMC). The results of all three approaches perfectly agree with each other as well as with the
theoretical result, Eq. (\ref{eq:alpha_heat}). For instance $T_k\sim k^{11/9}$ for $\gamma=1$ and  $T_k\sim
k^{17/9}$ for $\gamma=2$, which corresponds to the case of $\gamma \geq 2/3$ [see Eq. (\ref{eq:alpha_heat})].
Similarly, $T_k\sim k^{5/6}$ for $\gamma=0.5$, corresponding to $-1/3 \le \gamma <2/3$, see Fig. \ref{GAG1}a. All
these temperature distributions are the same as for the case of a velocity-independent restitution coefficient, with
the same driving coefficient $\gamma$.

Interestingly, for the equal distribution of the external energy supply for all species ($\gamma=0$) the energy
equipartition does not hold for particles of different sizes: $T_k\sim k^{1/3}$ which is confirmed both by the
scaling prediction and Monte Carlo simulations (Fig.~\ref{GAG1}b). The reason is that the losses of kinetic energy in collisions of smaller particles is larger than of bigger ones. In order to compensate this effect the input of the
external energy should be larger for smaller particles, which can be observed for negative values of $\gamma$. For the
system of viscoelastic particles the equipartition of energy takes place for $\gamma\simeq -0.47$ (Fig.~\ref{GAG1}b),
which is slightly different from the value $\gamma\simeq -0.33$, predicted by the scaling approach for granular systems
with constant restitution coefficient. This resembles the mimicry effect found in the binary mixtures of granular
particles \cite{mimic1,mimic2}. For negative values of $\gamma$ with larger absolute values the equipartition breaks
again, but the mass-dependence of the granular temperatures becomes inverse: The granular temperature of larger
particles becomes smaller \cite{zippelius}. This is indeed observed for the force controlled driving
with $\gamma=-1$ (Fig.~\ref{GAG1}b). All these findings are confirmed by the DSMC results.

As we have demonstrated above, the viscoelastic collision model yields the same temperature distribution as the
simplified collision model of a constant $\varepsilon$. The same is true for all dissipative mechanisms and may be
formulated as a general theorem:

\noindent {\bf Theorem:} \emph{Distribution of partial granular temperatures in a driven granular mixture does not
depend on the dissipation mechanism of inelastic collisions provided the size distribution of particles is steep enough.
}

\begin{figure}\centerline{\includegraphics[width=0.6\textwidth]{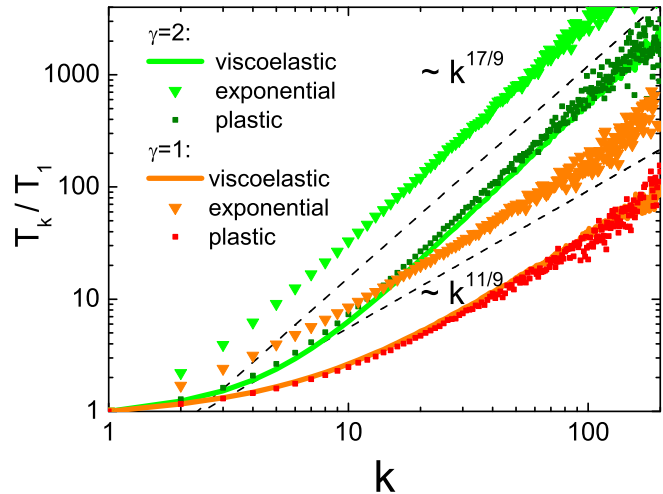}}
\caption{ Dependence of the granular temperatures $T_k$ on the reduced mass of particles $k=m_k/m_1$ for a heated
granular gas for different models of restitution coefficients: the viscoelastic model, elasto-plastic model with $V_{\rm yield}=1$
\cite{Thornton} and the exponential model with $\delta\sqrt{\frac{\rho}{Y}}=1$ \cite{LunSavage} obtained in the  DSMC simulations.The dashed lines indicate the slopes. }
\label{Gepsall}
\end{figure}

\noindent The proof of the theorem and exact formulation of the applicability conditions are given in the section
Methods. To illustrate the application of the general theorem we consider the temperature distribution in granular
mixtures with other mechanisms of the dissipative collisions -- the elasto-plastic model, described by Eq.
(\ref{eplastic}) \cite{Thornton} and the exponential model, described by Eq.~(\ref{eLun}) \cite{LunSavage}. The results
of DSMC simulation for granular mixtures with three different models of the restitution coefficient are shown in Fig.
\ref{Gepsall}. 

One can see from the figure that the distribution of granular temperatures demonstrates the same slope
for all mechanisms of the dissipative collisions. Moreover, for the case of the first-principle restitution coefficients
-- for viscoelastic and elasto-plastic dissipative mechanisms, not only the slopes of the distributions, but the distributions
themselves coincide. The discrepancy for small $k$ between the temperature distribution of the latter two models and
the phenomenological model from Ref.  \cite{LunSavage} stems possibly from the unphysically steep decay of
$\varepsilon$ with the impact velocity.

\section*{Conclusion}

We have studied kinetic properties of size-polydisperse granular mixtures where granular particles suffer pairwise
inelastic collisions. These are characterized by the restitution coefficients $\varepsilon$, quantifying the energy losses. We
consider two main mechanisms of the collisional dissipation, associated with the viscoelastic  and elasto-plastic
behavior of the particles material.  We used the first-principle expressions for restitution coefficients for these two
mechanisms, which implies the dependence of $\varepsilon$ on the impact velocity $v_{\rm imp}$, that is, $\varepsilon
=\varepsilon (v_{\rm imp})$. We also considered a  phenomenological exponential model for the impact-velocity dependent restitution
coefficient  $\varepsilon (v_{\rm imp})$ and a simplified model of a constant $\varepsilon$ that does not depend on the
impact velocity. We analyzed both cases  of force free and driven systems. We derived a system of equations that
describe the evolution of granular temperatures of different species in the mixture and the according cooling
coefficients $\xi_{ij}$ for the case of viscoelastic particles. We have also introduced the notion of the effective
impact-velocity independent restitution coefficient $\left \langle \varepsilon_{ij} \right \rangle$, which
significantly simplifies the analysis and demonstrated its adequacy and efficiency. We solved numerically the equations
for the granular temperatures $T_k$, where $k$ specifies the granular species and found the temperature distribution
$T_k$. The theoretical studies have been accompanied by the numerical modeling of the system with the direct
simulation Monte Carlo (DSMC). We observed an excellent agreement between the theoretical predictions and the numerical
results. Two main conclusions follow from our work: (i) The simplified model of a constant, velocity-independent
restitution coefficient fails qualitatively to describe the evolution of a granular mixture and the granular
temperature distribution in a homogeneous cooling state (HCS). (ii) Temperature distribution in driven granular
mixtures settles to a universal power-law distribution $T_k \sim k^{\alpha}$ with the exponent $\alpha$ that does not
depend on the dissipation mechanism of inelastic collisions. Moreover, the distribution of temperatures, obtained for
the two main dissipation mechanisms -- of viscoelastic and plastic energy losses, demonstrate not only coincidence of
the slopes but the overall coincidence. This kinetic law is applicable for granular mixtures with a steep size
distribution and may be formulated as a general theorem with a precise formulation of the conditions.

The reported results imply serious consequences for the overall kinetic behavior of granular materials -- the agitated
granular mixtures of very different nature may behave similarly. This result is important for fundamental science, as it helps to understand  the kinetic properties of granular mixtures and perform an adequate modelling, as well as for numerous practical applications.  As an immediate practical consequence of our findings, is the possibility of experimental modelling (imitation) of natural processes in a lab: One needs to care only about the size distribution of particles and the driving law, but not about particles' material.  Moreover, even if in the course of system  evolution, the mechanism of the dissipative collisions changes (e.g. from elasto-plastic to viscoelastic) it will not affect the temperature distribution.

\section*{Methods}
\subsection*{Derivation detail for the cooling coefficients, their sum and effective restitution coefficient}
Here we present the derivation detail for obtaining $\xi_{ki}$, $\sum_i \xi_{ki}$ and $\left< \varepsilon_{ik} \right>$.\\

\textit{Cooling coefficients} 
Due to rather small deviation of the velocity distribution function from the Maxwellian, we assume for this function the Maxwellian form, that is,
\begin{equation}
f_k(\textbf{v}_k)=n_k\left(\frac{m_k}{2\pi T_k}\right)^{3/2}\exp\left(-\frac{m_kv_k^2}{2T_k}\right)
\end{equation}
In order to compute the cooling rate we multiply the Boltzmann equation (Eq.~\ref{eq:BEgen}) by $m_kv_k^2/2$ and integrate it with respect to $v_k$ 
\begin{equation}
\frac{\partial\left\langle m_kv_k^2/2 \right\rangle}{\partial t}=\int d\textbf{v}_k \frac{m_kv_k^2}{2}\frac{\partial f}{\partial t}=\sum_i\int d\textbf{v}_k \frac{m_kv_k^2}{2} I_{ki}^{\rm coll}
\end{equation}
Using Eq.~(\ref{II}) for the collisional integral we write, 
\begin{eqnarray}\label{Iprop}
\int d\textbf{v}_k \frac{m_kv_k^2}{2} I_{ki}^{\rm coll}=\sigma_{ki}^2\int d\textbf{v}_k d\textbf{v}_i d\textbf{e} \Theta\left(-\textbf{v}_{ki}\cdot\textbf{e}\right)\left|\textbf{v}_{ki}\cdot\textbf{e}\right|\chi f(\textbf{v}^{\prime\prime}_k,t)f(\textbf{v}^{\prime\prime}_i,t)\frac{m_kv_k^2}{2}-\\
-\sigma_{ki}^2\int d\textbf{v}_k d\textbf{v}_i d\textbf{e} \Theta\left(-\textbf{v}_{ki}\cdot\textbf{e}\right)\left|\textbf{v}_{ki}\cdot\textbf{e}\right|f(\textbf{v}_k,t)f(\textbf{v}_i,t)\frac{m_kv_k^2}{2} \nonumber.
\end{eqnarray}
Noticing that \cite{book}
\begin{equation}
\chi \left|\textbf{v}_{ki}\cdot\textbf{e}\right| d\textbf{v}_k d\textbf{v}_i = \left|\textbf{v}_{ki}^{\prime\prime}\cdot\textbf{e}\right| d\textbf{v}_k^{\prime\prime} d\textbf{v}_i^{\prime\prime},
\end{equation}
we recast the first integral in the r.h.s.  of \eqref{Iprop} into the form
\begin{equation}
\sigma_{ki}^2\int d\textbf{v}_k^{\prime\prime} d\textbf{v}_i^{\prime\prime} d\textbf{e} \Theta\left(-\textbf{v}_{ki}^{\prime\prime}\cdot\textbf{e}\right)\left|\textbf{v}_{ki}^{\prime\prime}\cdot\textbf{e}\right| f(\textbf{v}^{\prime\prime}_k,t)f(\textbf{v}^{\prime\prime}_i,t)\frac{m_kv_k^2}{2}
\end{equation}
Since the pre-collision velocities $v_k^{\prime\prime}$ and $v_i^{\prime\prime}$ are related to $v_k$ and $v_i$ in the same way as $v_k$ and $v_i$ to post-collision velocities $v_k^{\prime}$ and $v_i^{\prime}$, this integral may 
be rewritten as 
\begin{equation}
\sigma_{ki}^2\int d\textbf{v}_k d\textbf{v}_i d\textbf{e} \Theta\left(-\textbf{v}_{ki}\cdot\textbf{e}\right)\left|\textbf{v}_{ki}\cdot\textbf{e}\right| f(\textbf{v}_k,t)f(\textbf{v}_i,t)\frac{m_kv_k^{\prime 2}}{2}
\end{equation}
which finally yields, 
\begin{equation}
\frac{\partial\left\langle m_kv_k^2/2 \right\rangle}{\partial t}=\sum_i\sigma_{ki}^2\int d\textbf{v}_k d\textbf{v}_i d\textbf{e} \Theta\left(-\textbf{v}_{ki}\cdot\textbf{e}\right)\left|\textbf{v}_{ki}\cdot\textbf{e}\right| f(\textbf{v}_k,t)f(\textbf{v}_i,t)\Delta E_k
\end{equation}
where $\Delta E_k=m_k\textbf{v}_k^{\prime \,2}/2-m_k\textbf{v}_k^2/2$ is the difference of energy of a particle of mass $m_k$ after and before a collision. Let us introduce the center of mass velocity 
\begin{equation}
\textbf{V}=\frac{m_k\textbf{v}_k+m_i\textbf{v}_i}{m_k+m_i}.
\end{equation}
Using the collision rules, Eqs. (\ref{v1v2}), one can derive $\Delta E_k$:
\begin{equation}
\Delta E_k=-m_{\rm eff}\left(\varepsilon_{ki}+1\right)\left(\textbf{v}_{ki}\cdot\textbf{e}\right)\left(\textbf{V}\cdot\textbf{e}\right)+\frac{1}{2}\frac{m_{\rm eff}^2}{m_k}\left(\textbf{v}_{ki}\cdot\textbf{e}\right)^2\left(\varepsilon_{ki}^2-1\right)
\end{equation}
where the restitution coefficient $\varepsilon_{ki}$ is given by Eq. (\ref{epsx}) with the following elastic constant
$\kappa$,
\begin{equation}
\kappa= \frac{1}{\sqrt{2}} \left (\frac{3}{2}\right )^{3/2}\frac{Y}{1-{\nu}^2}\left(\frac{6}{\pi\rho
m_1^2}\right)^{\frac13},
\end{equation}
which is a function of the Young's modulus $Y$, Poisson ratio $\nu$, monomer mass  $m_1$ and the material density of
particles $\rho$ \cite{rpbs99,bshp96,book}. The dissipative constant $A$ quantifies the viscous properties of the particles'
material \cite{bshp96,goldobin}:
\begin{equation}
\label{A} A =
\frac{1}{Y}\frac{\left(1+\nu\right)}{\left(1-\nu\right)}\left(\frac43\eta_1\left(1-\nu+\nu^2\right)+\eta_2\left(1-2\nu\right)^2\right)
\end{equation}
where ${\eta}_1$ and ${\eta}_2$ are the viscosity coefficients.
Let us introduce the variable 
\begin{equation}
\textbf{b}=\textbf{V}+\textbf{v}_{ki}m_{\rm eff} \frac{T_i-T_k}{m_k T_i+m_i T_k}
\end{equation}
The change of energy attains the form:
\begin{equation}\label{deltaEk}
\Delta E_k=m_{\rm eff}\left(\varepsilon_{ki}+1\right)\left(\textbf{v}_{ki}\cdot\textbf{e}\right)\left(\textbf{b}\cdot\textbf{e}\right)+\left(\textbf{v}_{ki}\cdot\textbf{e}\right)^2\frac{m_{\rm eff}^2}{m_k}\left(1+\varepsilon_{ki}\right)\left(\frac{1+\varepsilon_{ki}}{2}-\frac{T_k\left(m_k+m_i\right)}{T_km_i+T_im_k}\right)
\end{equation}
and the whole integral reads:
\begin{equation}
\frac{dT_k}{dt}=-T_k\sum_i\xi_{ki}
\end{equation}
with
\begin{equation}
\xi_{ki}=-\frac{1}{3}\sigma_{ki}^2n_i\left(\frac{m_k}{2\pi T_k}\right)^{3/2}\left(\frac{m_i}{2\pi T_i}\right)^{3/2}\int d\textbf{v}_{ki}d\textbf{b} d\textbf{e} \Theta\left(-\textbf{v}_{ki}\cdot\textbf{e}\right)\left|\textbf{v}_{ki}\cdot\textbf{e}\right|\exp\left(-\frac{m_km_i}{2\left(T_km_i+T_im_k\right)}v_{ki}^2-\frac{1}{2}\left(\frac{m_k}{T_k}+\frac{m_i}{T_i}\right)\textbf{b}^2\right)\frac{\Delta E_k}{T_k}\,.\label{integ}
\end{equation}
Integration that refers to the first term of $\Delta E_k$ [see  Eq.~(\ref{deltaEk}] yields zero. After the integration over $b$ we get:
\begin{equation}
\int_0^{\infty}d\textbf{b}\exp\left(-\frac{1}{2}\left(\frac{m_k}{T_k}+\frac{m_i}{T_i}\right)\textbf{b}^2\right)=\left(\frac{2\pi T_iT_k}{T_km_i+T_im_k}\right)^{3/2}.
\end{equation}
And Eq.~(\ref{integ}) can be now presented as a sum over the following type of integrals:
\begin{equation}
\int d\textbf{e}d\textbf{v}_{ki}\Theta\left(-\textbf{v}_{ki}\cdot\textbf{e}\right)\left|\textbf{v}_{ki}\cdot\textbf{e}\right|\left(-\textbf{v}_{ki}\cdot\textbf{e}\right)^{2+n/20}\exp\left(-Rv_{ki}^2\right)=4\pi^2\frac{10}{40+n}R^{-3-\frac{n}{20}}\Gamma\left(3+\frac{n}{20}\right)
\end{equation}
with $R=m_km_i/\left(2\left(T_km_i+T_im_k\right)\right)$. Collecting all terms together, one gets Eq.~(\ref{xikvisc}).\\

\textit{Effective restitution coefficient} 
The derivation of the effective restitution coefficient, Eq.~(\ref{epseff}), may be performed analogously. Introducing Eq.~(\ref{epsx}) into Eq.~(\ref{colav}), we get
\begin{equation}
\label{epsx} \varepsilon_{ki} = 1 +\frac{1}{\left\langle  v_n\right\rangle } \sum^{20}_{j=1} h_j \left( A \kappa_{ki}^{2/5} \right)^{j/2} I_{ki}^j
\end{equation}
with
\begin{eqnarray}
\label{Ikij} 
&&I_{ki}^j=\int d\textbf{v}_k d\textbf{v}_i d\textbf{e} \Theta\left(-\textbf{v}_{ki}\cdot\textbf{e}\right)\left|\textbf{v}_{ki}\cdot\textbf{e}\right|^{\frac{10+j}{10}} f(\textbf{v}_k,t)f(\textbf{v}_i,t)=\\\nonumber
&&=\left(\frac{2\left(T_km_i+T_im_k\right)}{\pi m_km_i}\right)^{-3/2}\int d\textbf{e}d\textbf{v}_{12}e^{-Rv_{12}^2}\left|\textbf{v}_{ki}\cdot\textbf{e}\right|^{\frac{10+j}{10}}=2\sqrt{2\pi}\left(\frac{T_k}{m_k}+\frac{T_i}{m_i}\right)^{1/2}\frac{20}{20+j} \Gamma
\left(2+\frac{j}{20} \right)2^{\frac{j}{20}}\left(\frac{T_k}{m_k}+\frac{T_i}{m_i}\right)^{j/20}.
\end{eqnarray}\\

\textit{The sum of cooling rates} 
To compute the sum $\sum_i \xi_{ki}$ in Eq. (\ref{sys}) we analyze the structure of the r.h.s. of Eq. (\ref{int0}) for
$i  \ll k$, starting from the first term. For $\alpha \geq 1$ the leading term depends on $i$ and $k$ as
$k^{\alpha/2-5/6} i$ and  for $\alpha < 1$ as $k^{-1/3} i^{(\alpha+1)/2}$. The next terms with $2 \leq n \leq 20$ scale as
$A^{n/2}i^{-n/6}k^{(\alpha-1)n/20}$ for $\alpha \geq  1$ and as $A^{n/2}i^{-(1-\alpha)n/20 -1/6}$ for $\alpha < 1$. For small $A$ [which is the condition of the validity of (\ref{epsx})] these terms  may be neglected for $n \geq 2$ as compared to the
first term, which yields Eq. (\ref{eq:xi_gen}).

\subsection*{The rigorous prove of the universality of the temperature distribution for all dissipative mechanisms}
With mild assumptions we can prove that temperature distribution tends to the scaling form $k^{\alpha}$, where $\alpha$
does not depend on a particular model of the effective restitution coefficient. Note that effective restitution
coefficient exists for every non-negative, continuous and
bounded $\varepsilon(v)$ and can be found using Eq. (\ref{colav}), although one can possibly get different values for the effective restitution
coefficient and the effective squared restitution coefficient, which we denote respectively as $\left\langle {{\varepsilon _{ki}}} \right\rangle$ and $\left\langle {{\varepsilon _{ki}^2}} \right\rangle$ (Here we do not consider the case of very soft particles or
nano-particles, which may possess a negative restitution coefficient \cite{Saitoh,PoeshelNeGEps}).
We present a proof for the Maxwell distribution, but the same techniques can be used in other cases. For simplicity we assume that $m_1 = 1$ and
$g_{2}(\sigma_{ki}) = 1$ (the generalization is straightforward), and rewrite (\ref{xikconst}) as
\begin{align}
\xi_{ki}(t) & = \frac{8}{3}\sqrt{2\pi}n_i\left(  i^{1/3} + k^{1/3} \right)^{2} \left(T_k/k+T_i/i \right)^{1/2}
\nonumber
\\
& \times \left( \frac{i}{i+k}\left[1-\frac{i}{2(i+k)}\left(1+\left\langle \varepsilon_{ki}\right\rangle \right) \left( 1 + \frac{T_i k}{T_k i} \right) \right] \right. \label{eq:xi} \\
& \left. + \frac{i}{i+k}\left[\left\langle \varepsilon_{ki}\right\rangle-\frac{i}{2(i+k)}\left(\left\langle
\varepsilon_{ki}\right\rangle + \left\langle \varepsilon_{ki}^2\right\rangle \right) \left( 1 + \frac{T_i k}{T_k i}
\right) \right] \right). \nonumber
\end{align}

First we check the convergence of $ \sum\limits_{i = 1}^\infty  {\left| {{\xi _{ki}}} \right|}$. Estimating $\left|
{{\xi _{ki}}} \right|$ from (\ref{eq:xi}) we find that for $i \geqslant k$
\begin{equation}
\sum\limits_{i = k}^\infty  {\left| {{\xi _{ki}}} \right|}  \leqslant {\rm const} \cdot \sum\limits_{i = k}^\infty  n_i
{{i^{2/3}}\sqrt {1 + \frac{{{T_i}}}{i}} \sqrt {1 + \frac{{{T_k}}}{k}} \left( {1 + \frac{{{T_i}k}}{{{T_k}i}}} \right)}.
\end{equation}
Therefore, if $T_i$ increases with $i$ slower than $i$,  we have the convergence if $\sum\limits_{i = 1}^\infty
{{n_i}{i^{2/3}}}  < \infty$. In the opposite
case, when $T_i$ increases with $i$ faster than $i$, the series converges if 
\begin{equation}\label{cond-1}
\sum\limits_{i = 1}^\infty  {{n_i}{i^{ - 5/6}}T_i^{3/2}}  < \infty.
\end{equation}

Let us impose even more restrictions by the condition
\begin{equation}\label{eq:aftk}
\frac{{{T_k}\sum\limits_{i = k}^\infty  {\left| {{\xi _{ki}}} \right|} }}{{{k^\gamma }}}\mathop  \to \limits^{k \to
\infty } 0,
\end{equation}
which holds true, if
\begin{equation}\label{cond-2}
\frac{1}{{{k^\gamma }}}\sum\limits_{i = k}^\infty  {{n_i}{i^{2/3}}(T_k + T_i \frac{k}{i})\sqrt {1 + \frac{{{T_i}}}{i}}
} \mathop  \to \limits^{k \to \infty } 0.
\end{equation}
Consider now a partial sum from $i = i_0$ to $k$, which satisfies,
\[
\sum\limits_{i = {i_0}}^k {\left| {{\xi _{ki}}} \right|}  \leqslant \sum\limits_{i = {i_0}}^k {{n_i}i{k^{ - 1/3}}\sqrt
{1 + \frac{{{T_k}}}{k}} }
\]
and impose the condition
\begin{equation}\label{eq:befk}
\forall \epsilon  > 0\quad\exists {i_0}\quad\forall k \geqslant {i_0} \quad \frac{{{T_k}\sum\limits_{i = {i_0}}^k
{\left| {{\xi _{ki}}} \right|} }}{{{k^\gamma }}} < \epsilon,
\end{equation}
which is true if $\sum\limits_{i = 1}^\infty  {{n_i}i}  < \infty$,
and $T_k = O(k^{\alpha})$ for $\alpha$ from (\ref{eq:alpha_heat}). Combining conditions (\ref{eq:aftk}) and (\ref{eq:befk}) we observe that

\begin{equation}
  \forall \epsilon  > 0\quad\forall k \geqslant {i_0} \quad
  \frac{{{T_k}\sum\limits_{i = {i_0}}^\infty  {\left| {{\xi _{ik}}} \right|} }}{{{k^\gamma }}} < \epsilon,
\end{equation}
which essentially means  that the full series may be replaced by its first $i_0$ terms with any desired accuracy $\epsilon$
 ($i_0$ certainly depends on $\epsilon$). Hence for $k \gg i_0$ we have the same asymptotics for $T_k$, obtained for
 the incomplete sum, as the one obtained for the whole series. These differ only by the factor  $1 + \epsilon$,
 converging to $1$ as $i_0$ and $k$ increase  (hereinafter it is implied that  $\epsilon$ may be taken arbitrarily
 small).

Now we illustrate that for $k \gg i_0$ the terms of $\xi_{ki}$ (\ref{eq:xi}) do converge. We have

\[\begin{gathered}  {\left( {{i^{1/3}} + {k^{1/3}}} \right)^2} = \left( {1 + \epsilon } \right){k^{2/3}} \hfill
\\  \sqrt {\frac{{{T_i}}}{i} + \frac{{{T_k}}}{k}}  = \left\{ \begin{gathered}  \left( {1 + \epsilon } \right)
\sqrt {\frac{{{T_i}}}{i}} ,\quad {\text{if}}\;{T_k}/k \to 0 \hfill \\  \left( {1 + \epsilon } \right)\sqrt {{\rm const}
+ \frac{{{T_k}}}{k}} ,\quad {\text{otherwise}} \hfill \\ \end{gathered}  \right. \hfill \\  \frac{i}{{i + k}} = \left(
{1 + \epsilon } \right)\frac{1}{k} \hfill \\  \frac{i}{{i + k}} \cdot \frac{{{T_i}k}}{{{T_k}i}} \to 0\quad
{\text{if}}\;{T_k} \to \infty  \hfill \\ \end{gathered}. \]
The last condition means that the terms with the negative sign in (\ref{eq:xi}) disappear. Then we are left with
\begin{equation}\label{eq:xilast}
{\xi_{ki}}\left( t \right) = {\rm const} \cdot \left( {1 + \epsilon } \right){k^{ - 1/3}}\sqrt {\frac{{{T_i}}}{i} +
\frac{{{T_k}}}{k}} \left( {1 + \left\langle {{\varepsilon _{ki}}} \right\rangle } \right), \qquad k \gg 1 .
\end{equation}
If $\left\langle {{\varepsilon _{ki}}} \right\rangle$ is continuous and has a limit for $k \to \infty$, then $1 +
\left\langle {{\varepsilon _{ki}}} \right\rangle$ is also a a value (that depends only on $i$) between $1$ and $2$ up
to the factor $(1 + \epsilon)$. Solving then the stationary equation (\ref{Tsteady})  for $T_k$, 

\begin{equation} \label{eq:41}
T_k \sum_i \xi_{ki} =
\Gamma_1k^{\gamma}    
\end{equation}
 with $\xi_{ki}$ from \eqref{eq:xilast} leads to
\[
  T_k = {\rm const} \cdot (1 + \epsilon) k^{\alpha}
\]
with  $\alpha$ from (\ref{eq:alpha_heat}), which proves the asymptotics. 

Note that the  special case of $\alpha=1$  is to be treated with an additional care. In this case the limit of $\left\langle {{\varepsilon _{ik}}} \right\rangle$ for $k \to \infty$ may depend on the limit of $T_k/k^{\alpha}$; Eq. \eqref{eq:41} for $T_k$, with $\xi_{ki}$ from \eqref{eq:xilast}, turns into an equation 
for an implicit function $f(x, \epsilon)$ with  $x = T_k/k^{\alpha}$. In this particular case one also needs to check that $\xi_{ki}$, given by Eq.  (\ref{eq:xilast}) is a monotonous function of $x = T_k/k^{\alpha}$. Otherwise, multiple solutions of Eq. \eqref{eq:41} for $\mathop {\lim }\limits_{k \to \infty } \frac{{{T_k}}}{{{k^\alpha }}}$ may exist. In all the addressed above cases,  the quantity  $\left\langle {{\varepsilon _{ik}}} \right\rangle$ decreases slower than $1 - c |\vec v|$; hence the monotony is guaranteed, so that $\mathop {\lim }\limits_{k \to \infty } \frac{{{T_k}}}{{{k^\alpha }}} = {\rm const} > 0$ even for $\alpha = 1$.

Substituting $T_k \sim k^{\alpha}$ into (\ref{cond-1}) and (\ref{cond-2}) we see that the following conditions are sufficient for the convergence for  the obtained asymptotics:
\begin{enumerate}
\item $n_i = {\cal O} \left(1/i^{\max(2, \gamma+1)+\delta} \right), \quad {\rm with} \quad \delta > 0$
\item $\gamma > -1/3$
\end{enumerate}
Surprisingly, there are no conditions imposed on the effective restitution coefficients $\left\langle \varepsilon_{ik}
\right\rangle$, except for the natural ones -- continuity and existence of some limit for high speeds, which are
obviously satisfied due to physical reasons. It should be also noted however that the power-law asymptotics is valid
only for large $k$ and is not expected for $k \sim 1$. In our simulations we used $n_k \sim  k^{-3}$, which means  that the monomers dominate in the system and the convergence to the asymptotics is already visible for rather small $k$.

\subsection*{Direct Simulation Monte Carlo}
To study the granular mixtures we implement the method of Direct Simulation Monte Carlo, which is widely used in
investigations of granular systems, especially for granular gases, where it provides very accurate results  \cite{Sant2000,NC}. It is based on the solution of the Boltzmann equation for space uniform systems by  stochastic methods \cite{Sant2000,TS}.  Generally, we follow the standard procedure \cite{TS}, which has been
adopted  accordingly for the granular mixture. Here we give the simulation detail and the main ideas of the approach.
We used $2.7 \times 10^8$ monomers, so that for the size distribution $n_k \sim 1/k^{3}$ addressed here, one has ${N_k} = \left\lfloor {{N_1}/{k^3}} \right\rfloor  = \left\lfloor
{2.7 \cdot {{10}^8}/{k^3}} \right\rfloor $ $k$-mers. The maximum size in our simulations was $k_{\max} = 300$. We use the following parameters: the mass of monomer $m_1=0.01$, its diameter, $\sigma_1=1$, the number  density of monomers $n_1=0.1$ and the initial temperature $T_0=1$. These parameters correspond to a dilute  granular mixture with the packing fraction of monomers $\phi = (\pi/6) n_1 \sigma_1^3 =0.0524$, which guarantees the accuracy of the Boltzmann equation \cite{book}. Note that any sufficiently small packing fraction leads to the same distribution, since scaling number density leads to the same changes in the equations and simulations as changing the time scale by the same amount.

A stochastic thermostat has been implemented through the random change of velocities of all particles every $Nh$
collisions, where $N$ is the total number of particles and $h$ is a parameter, which should be much less than one so
that the thermostat affects any particular particle several times between its collisions. Here we choose $h = 0.1$.

The velocity components of each particle changes each $Nh$ collisions:
\begin{align}
  {v_x} & := {v_x} + r_1 \sqrt {{\Gamma _i \Delta t}/{m_i}}, \nonumber \\
  {v_y} & := {v_y} + r_2 \sqrt {{\Gamma _i \Delta t}/{m_i}}, \nonumber \\
  {v_z} & := {v_z} + r_3 \sqrt {{\Gamma _i \Delta t}/{m_i}}, \nonumber
\end{align}
where $\Delta t$ is the time passed after previous velocity change due to the thermostat, and $r_{1,2,3}$ are random variables from a normal distribution with zero mean and unit standard deviation, that is,
$r_{1,2,3} \in N(0,1)$.

We consider only binary collisions of the particles and neglect possible triple and higher order collisions. Colliding particles are
chosen in two steps:
\begin{enumerate}
\item We choose the sizes of the colliding particles.
\item We choose the speeds of the colliding particles.
\end{enumerate}

Let us discuss each step in more detail.

1. We can estimate the number of candidate pairs $(ij)$ to collide during the
small time interval $\Delta t_{ij}$ as
\begin{equation}\label{eq:n_p}
N_p^{ij}  = \frac{\pi \left( \frac{\sigma_i + \sigma_j}{2} \right)^2 N_i N_j \left( |v_i|_{\max} + |v_j|_{\max} \right)}{V} \Delta t_{ij}
= \pi \frac{n_1}{N_1} N_i N_j \sigma_1^2 \left( \frac{i^{1/3} + j^{1/3}}{2} \right)^2 \left( |v_i|_{\max} + |v_j|_{\max} \right)
\Delta t_{ij},
\end{equation}
where $ (1/4) \pi (\sigma_i + \sigma_j)^2\left( |v_i|_{\max} + |v_j|_{\max} \right)\Delta t_{ij}$ gives the volume of
the collision cylinder and we use the definition of the number density, $n_i=N_i/V$ ($V$ is the system volume).  Let us introduce the matrix $\Lambda_{ij}$ of collision rates. They are equal to the inverse of the
average time of one collision between particles of sizes $i$ and $j$, that is, $\Lambda_{ij}= 
N_p^{ij} \Delta t_{ij}^{-1}$. Then we get from (\ref{eq:n_p}):
\[
  \Lambda_{ij} = \pi \frac{n_1}{N_1} N_i N_j\sigma_1^2 \left( \frac{i^{1/3} + j^{1/3}}{2} \right)^2 \left( |v_i|_{\max} + |v_j|_{\max} \right).
\]
The time between collisions and the sizes of colliding particles are chosen based on matrix $\| \Lambda_{ij} \|$. Namely,
the time between collisions is taken from the exponential distribution
\[
  \Delta t = - \ln ( \mathrm{rand} (0,1] ) / \sum\limits_{i,j = 1}^{m_{\max}} {{\Lambda_{ij}}}
\]
and the sizes $i$ and $j$ of the colliding particles are chosen with probability
\[
  p_{ij} = \frac{\Lambda_{ij}}{\sum\limits_{i,j = 1}^{m_{\max}} {{\Lambda_{ij}}}}.
\]
Note that matrix $\| \Lambda_{ij} \|$ has rank 6 and can be rapidly
recalculated using the low-rank technique (see e.g. \cite{Matrix}).\\

2. When the sizes $i$ and $j$ of the colliding particles are known, we can estimate the maximum value of the relative velocity of colliding particles according to
\[
  |v_{ij}|_{\max} \leqslant |v_i|_{\max} + |v_j|_{\max}.
\]
Then we use the standard technique to pick random particles and accept the collision if
\[
  | \textbf e (\textbf v_i - \textbf v_j)| > {\rm rand[0,1)} (|v_i|_{\max} + |v_j|_{\max}),
\]
where $\textbf e$ is a random unit vector. The values of $|v_i|_{\max}$ can be updated in ${\cal O} (\log N_i)$ steps by
keeping all velocities in a binary heap structure for each size.\\

To reduce the statistical noise of the simulation data for temperatures we apply the running averaging as follows. Let
at time $t_{\rm conv}$ the temperature of monomers ceases to change monotonously. This indicates that the system has
achieved its steady state and the stochastic noise becomes the primary source of errors. We then calculate the  temperatures every $N$ collisions in the time interval $(t_{\rm conv},\, 2t_{\rm conv})$ and take an average.



\begin{thebibliography}{99}

\bibitem{DryGranMed} Herrmann, H.~J., Hovi, J.-P., Luding, S.  \textit{Physics of Dry  Granular Media.} (Dordrecht: NATO ASI Series, Kluwer, 1998).
\bibitem{GranRev} Jaeger, H., Nagel, S., Behringer, R. Granular solids, liquids, and gases. \textit{Rev. Mod. Phys.} \textbf{68}, 1259-1273 (1996).
\bibitem{PhysGranMed} Hinrichsen, H., Wolf, D.~E. \textit{The Physics of Granular Media.} (Berlin: Wiley, 2004).
\bibitem{SanPowGr} Duran, J., Sands, Powders and Grains (Berlin: Springer-Verlag, 2000).
\bibitem{protodust}  Apai, D., Lauretta, D. S. \textit{Protoplanetary Dust. Astrophysical and Cosmochemical Perspectives.} (Cambridge University Press, 2010).
\bibitem{MoonSoil} Heiken, G.H.,  Vanniman, D.T. and French, B.M. \textit{Lunar Sourcebook.} (Cambridge University Press, 1991). 
\bibitem{mars} Almeida, M. P., Parteli, E. J. R., Andrade, J. S. Jr., Herrmann, H. J.
Giant saltation on Mars. \textit{Proc. Natl. Acad. Sci. USA} \textbf{105}, 6222–6226 (2008).
\bibitem{Dunes} Wiggs, G.F.S.,  Desert dune processes and dynamics. \textit{Progress in Physical Geography}  \textbf{25}, 53-79. (2001).
\bibitem{rings} Bridges, F. G., Hatzes, A., Lin, D. N. C. Structure, stability and evolution of Saturn's rings. \textit{Nature} \textbf{309}, 333-335 (1984).
\bibitem{pnas} Brilliantov, N., Krapivsky, P. L., Bodrova, A., Spahn, F., Hayakawa, H., Stadnichuk, V. and Schmidt,  J. Size distribution of particles in Saturn's rings from
aggregation and fragmentation \textit{Proc. Natl. Acad. Sci. USA} \textbf{112}, 9536-9541 (2015).
\bibitem{av1} Tai, Y.-C., Hutter, K., Gray, J. M. N. T. \textit{Geomorphological Fluid Mechanics, Lecture Notes in Physics} \textbf{582}, 339 (2001).
\bibitem{rpbs99}  Ramirez, R., P\"oschel, T., Brilliantov, N. and Schwager, T. Coefficient of restitution of colliding viscoelastic spheres. \textit{Phys. Rev. E} \textbf{60}, 4465-4472 (1999).
\bibitem{lev} Bodrova, A., Levchenko, D., Brilliantov, N.V. Universality of temperature distribution in granular gas mixtures with a steep particle size distribution. \textit{Europhys. Lett.} \textbf{106}, 14001 (2014).
\bibitem{book} Brilliantov, N. V. and P\"oschel, T. \textit{Kinetic theory of Granular Gases}. (Oxford: Oxford University Press, 2004)
\bibitem{brey}  Brey, J. J., Ruiz-Montero, M. J., Garcia-Rojo, R. and Dufty, J. W. Brownian motion in a granular gas. \textit{Phys. Rev. E} \textbf{60}, 7174-7181 (1999).
\bibitem{dufty}  Dufty, J. W. and Brey, J. J. Brownian motion in a granular fluid. \textit{New J. of Phys.} \textbf{7}, 20 (2005).
\bibitem{GarzoDuftyMixture}  Garzo, V. and Dufty, J. W. Homogeneous cooling state for a granular mixture. \textit{Phys. Rev. E} \textbf{60}, 5706-5713 (1999).
\bibitem{Hrenya}  Dahl, S. R., Hrenya, C. M., Garzo, V. and Dufty, J. W. Kinetic temperatures for a granular mixture. \textit{Phys. Rev. E} \textbf{66}, 041301 (2002).
\bibitem{wildman}  Wildman, R. D. and Parker, D. J. Coexistence of Two Granular Temperatures in Binary Vibrofluidized Beds. \textit{Phys. Rev. Lett} \textbf{88}, 064301 (2002). 
\bibitem{menon}  Feitosa, K. and Menon, N. Breakdown of Energy Equipartition in a 2D Binary Vibrated Granular Gas. \textit{Phys. Rev. Lett} \textbf{88}, 198301 (2002).
\bibitem{ringbook}  Schmidt, J., Ohtsuki, K., Rappaport, N., Salo, H. and Spahn, F. \textit{Dynamics of Saturn's Dense Rings.} In: Dougherty, M. K., Esposito, L. W., Krimigis, S. M. (Eds.) \textit{Saturn from Cassini-Huygens} (Springer) \textbf{413} (2009).
\bibitem{Ohtsuki1999}  Ohtsuki, K. Evolution of Particle Velocity Dispersion in a Circumplanetary Disk Due to Inelastic Collisions and Gravitational Interactions. \textit{Icarus} \textbf{137}, 152-177 (1999).
\bibitem{Ohtsuki2006}  Ohtsuki, K. Rotation rate and velocity dispersion of planetary ring particles with size distribution II. Numerical simulation for gravitating particles. \textit{Icarus} \textbf{183}, 384-395 (2006).
\bibitem{Salo1992b}  Salo, H. Numerical simulations of dense collisional systems: II. Extended distribution of particle sizes. \textit{Icarus} \textbf{96}, 85-106 (1992).
\bibitem{frank2004} Spahn, F., Albers, N.  Srem{\v c}evi{\'c}, M., Thornton, C. Kinetic description of coagulation and fragmentation in dilute granular particle ensembles. \textit{Europhys. Lett.}, \textbf{67}, 545–551 (2004).
\bibitem{h83}  Haff, P. Grain flow as a fluid-mechanical phenomenon. \textit{J. Fluid Mech.} \textbf{134}, 401-430 (1983).
\bibitem{gz93} Goldhirsch, I. and Zanetti, G. Clustering instability in dissipative gases. \textit{Phys. Rev. Lett.} \textbf{70}, 1619-1622 (1993). 
\bibitem{dp03} Das, S. and Puri, S. Pattern formation in the inhomogeneous cooling state of granular fluids. \textit{Europhys. Lett.} \textbf{61}, 749-755 (2003).
\bibitem{dppre03}  Das, S. and Puri, S. Kinetics of inhomogeneous cooling in granular fluids. \textit{Phys. Rev. E} \textbf{68}, 011302 (2003).
\bibitem{n03}  Nakanishi, H. Velocity distribution of inelastic granular gas in a homogeneous cooling state. \textit{Phys. Rev. E} \textbf{67}, 010301 (2003).
\bibitem{nebo79}  van Noije, T., Ernst, M., Brito, R. and Orza, J. Mesoscopic Theory of Granular Fluids. \textit{Phys. Rev. Lett.} \textbf{79}, 411-414 (1997).
\bibitem{neb98}  van Noije, T., Ernst, M. and Brito, R. Spatial correlations in compressible granular flows. \textit{Phys. Rev. E} \textbf{57}, R4891 (1998).
\bibitem{ap06}  Ahmad, S. and Puri, S. Velocity distributions in a freely evolving granular gas. \textit{Europhys. Lett.} \textbf{75}, 56-62 (2006).
\bibitem{ap07}  Ahmad, S. and Puri, S. Velocity distributions and aging in a cooling granular gas. \textit{Phys. Rev. E} \textbf{75}, 031302 (2007).
\bibitem{w60}  Goldsmit, W. \textit{The Theory and Physical Behavior of Colliding Solids} (Oxford: Oxford University Press, 2004)
\bibitem{bhd84}  Bridges, F., Hatzes, A. and Lin, D. Structure, stability and evolution of Saturn's rings. \textit{Nature} \textbf{309} 333-335 (1984).
\bibitem{kk87}  Kuwabara, G. and Kono, K. J. Restitution Coefficient in a Collision between TwoSpheres. \textit{Appl. Phys. Part 1} \textbf{26}, 1230-1233 (1987).
\bibitem{titt91}  Tanaka, T., Ishida, T. and Tsuji, Y. Trans. Direct Numerical Simulation of Granular Plug Flow in a Horizontal Pipe : the Case of Cohesionless Particles. \textit{Jap. Soc. Mech. Eng.} \textbf{57}, 456-463 (1991).
\bibitem{bshp96}  Brilliantov, N., Spahn, F., Hertzsch, J. and P\"oschel, T. Model for collisions in granular gases. \textit{Phys. Rev. E} \textbf{53}, 5382-5392 (1996).
\bibitem{mo97}  Morgado, W. and Oppenheim, I. Energy dissipation for quasielastic granular particle collisions. \textit{Phys. Rev. E} \textbf{55}, 1940-1945 (1997).
\bibitem{sp98}  Schwager, T. and P\"oschel, T. Coefficient of normal restitution of viscous particles and cooling rate of granular gases. \textit{Phys. Rev. E} \textbf{57}, 650-654 (1998).
\bibitem{delayed}  Schwager, T. and P\"oschel, T. Coefficient of restitution for viscoelastic spheres: The effect of delayed recovery. \textit{Phys. Rev. E} \textbf{78}, 051304 (2008).
\bibitem{Falcon1} Falcon, E., Laroche, C., Fauve, S. and Coste, C. Behavior of one inelastic ball bouncing repeatedly off the ground, 
\textit{Eur. Phys. J. B } \textbf{3}, 45-57 (1998).
\bibitem{Labous} Labous, L., Rosato, A.D. and Dave, R.N., Measurements of collisional properties of spheres using high-speed video analysis, \textit{Phys. Rev. E} \textbf{56}, 5717 (1997)
\bibitem{Falcon2} McNamara, S. and  Falcon, E. Simulations of dense granular gases without gravity with impact-velocity-dependent restitution coefficient, \textit{Powder Technology} \textbf{182},  232–240, (2008).
\bibitem{goldobin} Goldobin, D. S., Susloparov, E. A., Pimenova, A. V., Brilliantov, N. V. Collision of viscoelastic bodies: Rigorous derivation of dissipative force. \textit{Eur. Phys. J. E} \textbf{38}, 55 (2015).
\bibitem{Thornton} Thornton, C. and Ning, Z. A theoretical model for the stick/bounce behaviour of adhesive, elastic-plastic spheres. Powder Technology \textbf{99}, 154-162 (1998).
\bibitem{LunSavage} Lun, C. K. K. and Savage, S. B. The Effects of an Impact Velocity Dependent Coefficient of Restitution on Stresses Developed by Sheared Granular Materials. \textit{Acta Mechanica} \textbf{63}, 15-44 (1986).
\bibitem{NBTPSelfDif2000} Brilliantov, N.V. and Poeschel, T. Self-diffusion in granular gases, \textit{Phys. Rev. E} \textbf{61},  1716-1721  (2000). 
\bibitem{ClustersPRL2004} Brilliantov, N.V., Saluena, C., Schwager, T. and Poeschel, T. Transient structures in a granular gas, \textit{Phys. Rev. Lett.} \textbf{93},  134301 (2004). 
\bibitem{expgg}  Grasselli, Y., Bossis, G. and Goutallier, G. Velocity-dependent restitution coefficient and granular cooling in microgravity. \textit{Europhys. Lett.} \textbf{86}, 60007 (2009).
\bibitem{briltemp} Brilliantov, N. V. and P\"oschel, T. Velocity distribution in granular gases of viscoelastic particles. \textit{Phys. Rev. E} \textbf{61}, 5573-5587 (2000).
\bibitem{annaprl} Bodrova, A. S., Dubey, A. K., Puri, S. and Brilliantov, N. V. Intermediate Regimes in Granular Brownian Motion: Superdiffusion and Subdiffusion. \textit{Phys. Rev. Lett.} \textbf{109}, 178001 (2012). 
\bibitem{annapre}  Dubey, A. K., Bodrova, A., Puri, S. and Brilliantov, N. Velocity distribution function and effective restitution coefficient for a granular gas of viscoelastic particles. \textit{Phys. Rev. E} \textbf{87}, 062202 (2013).
\bibitem{WilliamsMacKintosh1996} Williams, D. R. M. and MacKintosh, F. C. Driven granular media in one dimension: Correlations and equation of state. \textit{Phys. Rev. E} \textbf{54}, R9 (1996).
\bibitem{Sant2000} Montanero, J. M. and Santos, A. Computer simulation of uniformly heated granular fluids. \textit{Gran. Mat.} \textbf{2}, 53-64 (2000).
\bibitem{vne98} van Noije, T. and Ernst, M. Velocity distributions in homogeneous granular fluids:
the free and the heated case. \textit{Gran. Mat.} \textbf{1}, 57-64 (1998).
\bibitem{zippelius} Uecker, H., Kranz, W. T., Aspelmeier T. and Zippelius A. Partitioning of energy in highly polydisperse granular gases. \textit{Phys. Rev. E} \textbf{80}, 041303 (2009).
\bibitem{Cafiero} Cafiero, R., Luding, S. and Herrmann, H. J. Two-Dimensional Granular Gas of Inelastic Spheres with Multiplicative Driving. \textit{Phys. Rev. Lett.} \textbf{84}, 6014-6017 (2000).
\bibitem{mimic1} Lasanta, A., Vega Reyes, F., Garzo, V., Santos, A. Intruders in disguise: Mimicry effect in granular gases. https://arxiv.org/abs/1903.10807.
\bibitem{mimic2} Megias, A., Santos, A. Driven and undriven states of multicomponent granular gases of inelastic and rough hard disks or spheres. https://arxiv.org/abs/1901.11307.
\bibitem{NC} Brilliantov, N., Poeschel, T, and Formella, A., Increasing temperature of cooling granular gases \textit{Nature Communication}, \textbf{9},  797 (2018).
\bibitem{TS} T. Poschel and T. Schwager, \textit{Computational Granular Dynamics}. (Springer - Berlin, Heidelberg, New York, 2005).

\bibitem{Saitoh} Saitoh, K., Bodrova, A., Hayakawa, H. and Brilliantov, N.V. Negative Normal Restitution Coefficient Found in Simulation of Nanocluster Collisions, \textit{Phys. Rev. Lett.}, \textbf{105}, 238001 (2010)
\bibitem{PoeshelNeGEps} Muller, P., Krengel, D. and Poschel, T.,  Negative coefficient of normal restitution, \textit{Phys. Rev. E} \textbf{85}, 041306 (2012).
\bibitem{Matrix} R. A. Horn, and C. R. Roger  \textit{Matrix
analysis}. Second edition. (Cambridge University Press,
2013).
\end{thebibliography}
\end{document}